\documentclass[final,5p,times]{elsarticle}
\usepackage[utf8]{inputenc}
\usepackage{hyperref}
\usepackage{float}
\usepackage{amsmath}
\usepackage{notes2bib}
\usepackage{physics}
\usepackage{tikz}
\usetikzlibrary{positioning}
\usetikzlibrary{arrows.meta}
\usetikzlibrary{shapes}
\usepackage{tabularx}
\usepackage[caption = false]{subfig}
\usepackage{bbold}
\usepackage{siunitx}

\usepackage{amssymb}

\newcounter{bla}

\journal{Computer Physics Communications}

\begin{document}

\begin{frontmatter}



\title{\textsc{tinie} -- a software package for electronic transport through two-dimensional cavities in a magnetic field}


\author[a]{R. Duda}
\author[a]{J. Keski-Rahkonen\corref{author}}
\author[a]{J. Solanpää}
\author[a]{E. Räsänen}

\cortext[author] {Corresponding author.\\\textit{E-mail address:} joonas.keski-rahkonen@tuni.fi}
\address[a]{Computational Physics Laboratory, Tampere University, P.O. Box 692, FI-33014 Tampere, Finland}

\begin{abstract}
Quantum transport has far-reaching applications in modern electronics as it enables the control of currents in nanoscale systems such as quantum dots. In this paper we introduce \textsc{tinie}: a state-of-the-art quantum transport simulation framework, which can efficiently perform first-principle calculations based on the Landauer-Büttiker formalism. The computational repertoire of \textsc{tinie} includes calculations of transmission, conductivity, and currents running through arbitrary multi-terminal two-dimensional transport devices, with additional tools that enable the computation of the local density of states. The generality of \textsc{tinie} ranges from wide-band approximation calculations to investigating systems subject to an external magnetic field. The future prospects of \textsc{tinie} include the simulation of, e.g., two-dimensional cavities, quantum dots, or molecular junctions. The package is written in Python 3.6, and its well-documented modular structure is designed with an intent to create a platform suited for continuous expansion and development. With \textsc{tinie} it is possible to obtain specific information about the effects of impurities and imperfections in quantum devices, particularly between ballistic and diffusive transport regimes.

\end{abstract}

\begin{keyword}
quantum transport; low-dimensional systems; quantum dots

\end{keyword}

\end{frontmatter}



{\bf PROGRAM SUMMARY}

\begin{small}
\noindent
{\em Program title:} \textsc{tinie}                                          \\
{\em Licensing provisions:} Boost Software License 1.0 \\
{\em Programming language:} Python 3.6                                \\
{\em Computer:} Tested on x86\_64 architectures. \\
{\em Operating systems:} Tested on Linux/Mac OS \\
{\em Parallelization:} Parallelized with \texttt{mpi4py}. \\
{\em Nature of the problem:}\\
Numerical calculation of the properties of a two-dimensional nanoscale electron transport system in a uniform magnetic field (zero or non-zero), specifically the currents running through the reservoirs (leads) coupled to a quantum dot (central region) and the corresponding transmission coefficients. \\
{\em Solution method:}\\
The problem solution is split into two stages. The first stage (\textsc{tinie\_prepare} stage) prepares the transport system data for the main transport calculation. This data comprise Hamiltonian matrices of the uncoupled reservoirs and quantum dot regions, their respective sets of eigenfunctions and the coupling matrices between the quantum dot and the reservoirs. The second stage (\textsc{tinie} stage) performs the transport calculation for the given system using the embedding self-energy technique. \\
{\em Restrictions:} The code is restricted to the non-interacting equilibrium transport problems.\\
{\em Unusual features:} The code is modular in structure, allowing for easy extension and introduction of different reservoir/quantum dot/coupling types. Additionally, \textsc{tinie} is compatible with systems in a non-zero uniform magnetic field.\\
{\em Additional comments:} The source code is available at \url{https://gitlab.com/compphys-public/tinie} and Python package in \url{https://pypi.org/project/tinie/}. An extensive documentation of the code functionality can be found in the \texttt{README.md} file accompanying the code. \\
{\em Running time:} Seconds to days, depending on the simulation and parallelization.
   \\
\end{small}

\section{Introduction}
\label{introduction}
\noindent
Quantum transport is one of the most common -- and at the same time most tedious -- concepts in condensed matter physics dating back to the beginning of mesoscopic physics when the first transport algorithms were developed~\cite{Landauer1956, Lee1981, Thouless1981, MacKinnon1985, Buttiker1988, Landauer1988, Buttiker1989}. Solving a quantum transport problem provides access to currents, conductances, density of states, and other key properties of nanosystems including novel structures such as topological insulators, for example.

In this paper we present \textsc{tinie}: a modern, versatile implementation of the Green's function method for solving the equilibrium quantum transport properties of a generic two-dimensional (2D) nanostructure. 2D systems are not only convenient theoretical models,
but they are also experimentally realizable in various settings including, e.g., semiconductor structures~\cite{Micolich2013, Davies_book, Bimberg_book, Nakamura_Harayma_book}, quantum Hall systems~\cite{von_Klitzing1980, von_Klitzing1986}, topological insulators~\cite{Hasan2010, Bernevig_book}, quantum dots~\cite{Davies_book, Bimberg_book, Nakamura_Harayma_book, Turton_book, Michler_book, Harrison_book, Tartakovskii_book, Blood_book, Bruchez_book}, graphene~\cite{Novoselov2004, Geim2007, Geim2009} and other single-layer atomic systems. Nevertheless, we point out that the core functionalities of our implementation can be directly extended to a three-dimensional systems.

Although the Green's function formalism (see, e.g., Refs.~\cite{Datta1997, Ventra2008, Stefanucci2013}) used in our program has been employed before~\cite{Kwant, Nextnano,Transiesta, Smeagol, Openmx}, \textsc{tinie} approaches the quantum transport problem from a different point of view by separating it into two parts: (i) the eigenvalue problem of a closed system and (ii) the transport in a connected system. The first task is outsourced to external packages which can be chosen optimally for a given eigenvalue problem. The transport part can then be performed by employing the presented software package \textsc{tinie}. 

By default, \textsc{tinie} is designed to be compatible with \textsc{itp2d} package~\cite{Itp2d} which is optimized for solving tens of thousands of eigenstates of the time-independent Schrödinger equation for an arbitrary external potential, allowing various experimentally relevant shapes for quantum dots, for example. The \textsc{itp2d} package utilizes the imaginary time propagation method~\cite{Aichinger2005, Chin2009,Janecek2008}, which is particularly suited for 2D problems with perpendicular magnetic fields due to the existence of an exact factorization of the exponential kinetic energy operator in a magnetic field~\cite{Aichinger2005, Janecek2008_II, Chin2010}. However, \textsc{tinie} is directly compatible with other eigenfunction solvers as well. In addition, it is straightforward to combine \textsc{tinie} with real-space electronic-structure methods based on density-functional theory, e.g., the \textsc{octopus} code package~\cite{octupus}.

After solving the given eigenvalue problem, the quantum transport properties of an open system can be determined by employing the versatile numerical environment given by \textsc{tinie}. As an input, the transport code only requires the eigenenergies of the closed system and the matrices describing the coupling between the considered system and the attached leads. A tool for computing these coupling matrices is included. In general, \textsc{tinie} provides a way to study equilibrium quantum transport in a multi-terminal system with an arbitrary lead configuration at zero and finite temperature, even in the presence of an external, homogeneous magnetic field. \textsc{tinie} is written in modern Python (version 3.6). Furthermore, it offers a modular platform that can be easily extended without sacrificing the speed or user-friendliness.

The structure of the paper is as follows: In Sec.~\ref{sec:theory}, we introduce the quantum transport scheme behind our implementation, which is described in Sec.~\ref{sec:implementation}. In Sec.~\ref{sec:tests}, we present numerical results of a few prototype systems simulated with \textsc{tinie}. We finish with a brief discussion and the summary of the paper in Sec.~\ref{sec:summary}.

\section{Theoretical background}
\label{sec:theory}
\noindent
In order to give a self-contained presentation, we consider first the conventional 
Landauer-B\"{uttiker} approach within the Green's function formalism, which forms the theoretical basis for the quantum transport routines in \textsc{tinie}. A reader who is already familiar with the formalism can advance directly to Sec.~3, where we describe the design of our implementation in detail.

\subsection{Landauer-B\"{uttiker} formalism}
\noindent
A generic framework for quantum transport is covered by a scattering formalism (see, e.g., Ref.~\cite{Datta1997}). Instead of describing states in a closed geometry, we consider the scattering of electrons in a finite system, or a quantum device, coupled to infinite leads. The formalism developed by Landauer~\cite{Landauer1956, Landauer1988} and later complemented by B\"{u}ttiker~\cite{Buttiker1988, Buttiker1989} provides an intuitive physical description for a current flowing in a nanoscale or mesoscopic structure. The system is composed of reservoirs acting as leads and a central part -- or conduction device -- that describes a molecule or a quantum well, for example. The electronic current is understood in terms of transmission probabilities for an electron traveling from one reservoir to another through the conducting device. In the steady-state regime, the measured current in a reservoir is the difference between the currents flowing in and out of the reservoir.

Derived from the time-dependent Schr\"{o}dinger equation~\cite{Caroli_1971_I, Caroli_1971_II}, the Landauer-B\"{u}ttiker formula gives a microscopic understanding for a tunneling current in a transport setup. The leads are initially uncoupled to the central conducting device, being in equilibrium at different chemical potentials. After the coupling between the leads and the central device is switched on, the Landauer-B\"{u}ttiker formula is recovered as the long-time limit $t \rightarrow \infty$ of the expectation value of the current operator. However, the idea of instantaneous attaching of the leads to the central device is experimentally unreasonable. Thus, an alternative approach has been presented in Ref.~\cite{PhysRevB.22.5887}. Here, the whole system is assumed to be already initially connected but in equilibrium at a unique chemical potential, which is driven out of equilibrium by an applied bias voltage. Even though the initial point of view is different in this approach, the same Landauer-B\"{u}ttiker formula is recovered. As was shown later in Ref.~\cite{PhysRevB.69.195318}, the initial preparation of the system does not affect the steady-state limit described by the Landauer-B\"{u}ttiker formula.

The formalism of Landauer and B\"{u}ttiker may also be rigorously derived from the microscopic theory based on the non-equilibrium Green's function formalism (see, e.g., Ref.~\cite{Stefanucci2013}). The interpretation of the system's properties in terms of Green's functions is therefore completely equivalent to the Schr\"{o}dinger equation. Furthermore, the Fisher-Lee relation~\cite{Lee1981} connects the non-equilibrium Green's function formalism to the mathematically equivalent wave function formulation of the scattering problem. When applied to quantum transport, the non-equilibrium Green's function method enables the calculation of currents in a multi-terminal system for \emph{all times}. The steady-state value of current agrees with the Mier-Wingreen formula~\cite{PhysRevLett.68.2512, PhysRevB.50.5528}, under the assumptions that initial correlations and initial-state dependencies are washed out in the limit $t \rightarrow \infty$, and, in the same limit, the invariance under time translations is reached. Restricting to the non-interacting case, the well-known Landauer-B\"{u}ttiker formula~\cite{Landauer1956, Buttiker1989} is again obtained for the steady-state current.

\subsection{Transport setup}
\label{ssec:systemsetup}
\begin{figure}[ht]
    \centering
    \includegraphics[scale=0.3]{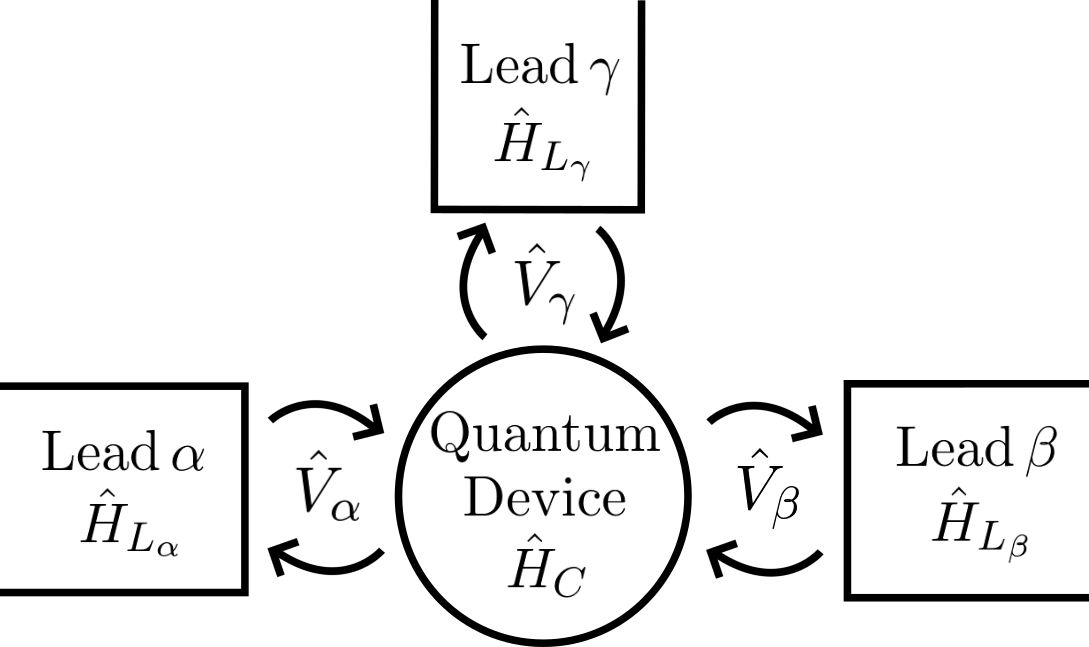}
    \caption{Typical setup of a multi-terminal transport system. In this case, three leads are coupled to the quantum device.}
    \label{fig:setup}
\end{figure}

\noindent
The Landauer-B\"{u}ttiker formalism offers a physically appealing first-principle approach to study steady-state currents in conducting quantum device such as a quantum dot~\cite{Unlu2012} (QD) or a molecule. We look at a specific setup of partitioning the system of interest into leads and a QD. The structure of a transport setup is illustrated for a three-terminal system in Fig.~\ref{fig:setup}.
In general, the Hamiltonian of the studied transport system can be divided into the block form of
\begin{displaymath}
\begin{pmatrix}
H_{L_{1}} & 0& 0 & \cdots & 0 & V_{1}\\
0 & H_{L_2} & 0 & \cdots & 0& V_{2}\\
\vdots &&  & \ddots & \vdots & \vdots\\
0 & 0 & 0 & \cdots & H_{L_N} & V_{N}\\
V_1^{\dagger} & V_2^{\dagger} & \cdots & & V_{N}^{\dagger} & H_C 
\end{pmatrix}.
\end{displaymath}
Here it is assumed that the leads $\alpha = 1, \dots, N$ are coupled only through the QD and the direct couplings between the leads are zero.

To determine the electric transport through the QD, we need three distinct elements appearing in the partition above: the Hamiltonian $\hat{H}_C$ of the QD, the Hamiltonians $\hat{H}_{{L}_{\alpha}}$ of all the attached leads, and the coupling $\hat{V}_{\alpha}$ between each lead and the QD. In particular, the eigenenergies $\{ E^{C} \}$ and the corresponding states $\{ \vert \psi_C \rangle \}$ of the isolated QD can be obtained by interfacing with a suitable eigenvalue solver, such as \textsc{itp2d} \cite{Luukko2013}. Secondly, the eigenfunctions $\{ \vert \psi_{L_{\alpha}} \rangle \}_{\alpha}$ and their energies are assumed to be either known analytically, or they can be solved numerically. For convenience, here we use the Latin indices for the individual states in the lead, whereas the Greek indices refer to the entire lead. Finally, the couplings defining the connection between the leads and the QD can be either be provided manually, or estimated numerically as discussed below in Sec. \ref{ssec:coupling}.  
 
\subsection{Embedding self-energy technique}
\label{ssec:eset}

\noindent
The blocks of a transport setup are combined together for quantum transport within the \emph{embedding self-energy technique}. Here the open system of a device connected to the reservoirs is mapped to a closed system, where the leads are taken into account as self-energy terms. Similar self-energy terms also capture the effects of the electron-phonon and electron-electron interactions, which are neglected here. The term \emph{embedding}~\cite{Datta1997, Ventra2008, Stefanucci2013} highlights the fact that the considered self-energies stem from the coupling between the central region and the lead environment. They can be viewed to result in an effective Hamiltonian arising from the interaction of the QD with the leads. 

The dynamics of an electron in a QD is encoded in retarded and advanced Green's functions~\cite{Stefanucci2013}:
\begin{equation}
\begin{aligned}
\label{eq:green}
\hat{G}^{R}(\omega) &= \left[ \omega\mathbb{1} - \hat{H}_C - \sum_{\alpha} \hat{\Sigma}^{R}_{\alpha}(\omega)\right]^{-1}, \\
\hat{G}^{A}(\omega) &= \left[ \omega\mathbb{1} - \hat{H}_C - \sum_{\alpha} \hat{\Sigma}^{A}_{\alpha}(\omega)\right]^{-1},
\end{aligned}
\end{equation}
where $\mathbb{1}$ is the identity operator. The couplings to the terminals are now taken into account by introducing the embedding retarded and advanced self-energies, which are defined, respectively, as
\begin{equation}
\begin{aligned}
\label{eq:selfenergy}
\hat{\Sigma}^{R}_{\alpha}(\omega) &= \hat{V}^{\dagger}_{\alpha} g_{\alpha}^{R}(\omega) \hat{V}_{\alpha} 
\quad \textrm{and} \quad
\hat{\Sigma}^{A}_{\alpha}(\omega) &= \hat{V}^{\dagger}_{\alpha} g_{\alpha}^A(\omega)\hat{V}_{\alpha},
\end{aligned}
\end{equation}
where $V_{\alpha}$ describes the coupling of the lead $\alpha$ to the QD, and $g_{\alpha}^R(\omega)$ and $g_{\alpha}^A(\omega)$ are the retarded and advanced Green's functions of the lead $\alpha$ at energy $\omega$, respectively (see, e.g., Ref.~\cite{Stefanucci2013}). In the terms of the lead Hamiltonian $H_{L_{\alpha}}$, these Green's functions are
\begin{displaymath}
\begin{aligned}
g_{\alpha}^R(\omega) &= \left[(\omega+\mathrm{i}\eta)\mathbb{1}-\hat{H}_{L_{\alpha}}\right]^{-1}\\
g_{\alpha}^A(\omega) &= \left[(\omega-\mathrm{i}\eta)\mathbb{1}-\hat{H}_{L_{\alpha}}\right]^{-1}.
\end{aligned}
\end{displaymath}
The positive infinitesimal $\eta$ accounts for the proper causal structure in the retarded and advanced Green's functions: the retarded Green's function is analytic in the upper-half of the complex plane, whereas the advanced function is analytic in the lower-half of the plane. We could have included an infinitesimal imaginary part in the Green's functions in Eq.~\eqref{eq:green} as well, but the self-energy stemming from the coupling to leads effectively gives rise to a \emph{finite} imaginary contribution that will swamp it.

For each lead $\alpha$, we associate a \emph{rate operator}~\cite{Stefanucci2013}, or a level-width function, which is given by the difference of the self-energies as 
\begin{equation}
\label{eq:rateoperator}
\hat{\Gamma}_{\alpha} = \mathrm{i} \left[ \hat{\Sigma}_{\alpha}^R(\omega) - \hat{\Sigma}_{\alpha}^A(\omega) \right].
\end{equation}
Thus, the imaginary part of the embedding self-energy functions~\eqref{eq:selfenergy} gives rise to the broadening of the energy levels. Intuitively, the real part of the embedding self-energy could be absorbed into the Hamiltonian of the device, therefore only shifting the poles of the Green's function, whereas the imaginary part of the embedding self-energy gives the width of the peaks. In a physical view, the coupling of the QD to the leads shifts its energy levels arising from the real part of the embedding self-energy. However, these levels have a finite life-time as an electron can escape from the QD to the leads, or vice versa, which is characterized by the rate-operators.

Similarly, one often encounters a difference of the retarded and advanced Green's functions, known as the \emph{spectral function}~\cite{Stefanucci2013},
\begin{displaymath}
\hat{A} = \mathrm{i} \left[ \hat{G}^R(\omega) - \hat{G}^A(\omega) \right].
\end{displaymath}
Since the spectral function describes the spectral density in the QD due to the lead self-energy, we may weigh the function by the occupation probability and integrate over the energy to obtain the non-equilibrium density matrix for the QD:
\begin{displaymath}
\hat{\rho} = \int \frac{\textrm{d} \omega}{2 \pi} f_{\textrm{FD}}(\omega - \mu) \hat{A}(\omega),
\end{displaymath}
where $\mu$ is chemical potential, and the Fermi-Dirac function with temperature $T$ is defined as
\begin{displaymath}
f_{\textrm{FD}}(\omega;T) = \left( e^{\omega/T} + 1 \right)^{-1}. 
\end{displaymath}

In the absence of interactions, the steady-state current can be expressed in the terms of the \emph{transmission} between two distinct leads $\alpha$ and $\beta$~\cite{Datta1997, Ventra2008, Stefanucci2013} as
\begin{equation}
\label{eq:transmission}
\mathcal{T}_{\alpha \beta} (\omega) = \textrm{Tr} \left[ \hat{G}^{R}(\omega) \hat{\Gamma}_{\beta}(\omega) \hat{G}^{A}(\omega) \hat{\Gamma}_{\alpha}(\omega) \right]  
\end{equation}
The transmission $\mathcal{T}_{\alpha \beta}$ is directly related to the probability for an electron of energy $\omega$ to be transmitted from the reservoir $\alpha$ to the reservoir $\beta$ via the QD. On the other hand, the backscattering transmission $\alpha = \beta$ is given by
\begin{equation}
\label{eq:backscattering}
\mathcal{T}_{\alpha \alpha} (\omega) = \textrm{Tr} \left[ \left( \mathbb{1} - i \hat{\Gamma}_{\alpha}(\omega) \hat{G}^{R}(\omega) \right) \left( \mathbb{1} - i \hat{\Gamma}_{\alpha}(\omega) \hat{G}^{R}(\omega) \right) \right].  
\end{equation}

The current contribution from the reservoir $\alpha$ to the reservoir $\beta$ is described by the \emph{partial current}
\begin{equation}
\label{eq:pcurrents}
\begin{split}
i_{\alpha\beta}=2\int\frac{1}{2\pi}\Big[&f_{\mathrm{FD}}(\omega-V_{\alpha}-\mu;T) \\ 
& -f_{\mathrm{FD}}(\omega-V_{\beta}-\mu;T)\Big]\mathcal{T}_{\alpha\beta}(\omega) \, \mathrm{d}\omega.
\end{split}\end{equation}
The factor of two in front of the integral stems from the spin degeneracy. The \emph{total current} is then determined as a sum over all partial currents for the particular lead:
\begin{displaymath}
I_{\alpha} = \sum_{\beta} i_{\alpha \beta}.
\end{displaymath}
It should be emphasized that the total current is unaffected by the backscattering~\eqref{eq:backscattering}, as the partial currents vanish for the case $\alpha =\beta$. In the linear response regime, the total current acquires an Ohmic form of
\begin{displaymath}
I_{\alpha} = \sum_{\beta} \mathcal{G}_{\alpha \beta} \left( V_{\alpha} - V_{\beta} \right),
\end{displaymath}
where the average conductance $\mathcal{G}_{\alpha \beta} $ is 
\begin{equation}
\label{eq:conductance}
    \mathcal{G}_{\alpha\beta}(\omega) = \frac{2}{2\pi}\int \, \mathcal{T}_{\alpha\beta}(\omega')\frac{1}{4T} \sech^{2}\left(\frac{\omega'-\omega}{2T}\right)\mathrm{d}\omega'.
\end{equation}
with $F_{\mathrm{TH}}(\omega)$ being the thermal broadening function determined as
\begin{displaymath}
F_{\mathrm{TH}}(\omega) = \frac{1}{4T} \textrm{sech}^2 \left( \frac{\omega}{2T} \right)
\end{displaymath}
at temperature $T$ (see, e.g., Ref.~\cite{Datta1997}). Furthermore, in the zero-temperature limit, the conventional Landauer-B{\"u}ttiker formula~\cite{Landauer1956, Buttiker1988, Landauer1988, Buttiker1989} is recovered:
\begin{equation}
    \lim_{T \rightarrow 0} \mathcal{G}_{\alpha\beta}(\omega) = \frac{2}{2\pi}\mathcal{T}_{\alpha\beta}(\omega).
\end{equation} 

\subsection{Gauge transformation for a magnetic field}
\label{ssec:bfield}
In case of a non-zero perpendicular and constant magnetic field $\mathbf{B}$, we must make sure that the gauge of the vector potential $\mathbf{A}$ is consistent with the various choices of the frame of reference of the central region and its accompanying reservoirs. 

The inclusion of an external magnetic field in \textsc{tinie} is one of its novelty factors. We assume that the vector potential of the central region is of the linear gauge form $\mathbf{A}=-By\hat{x}$. This choice for the gauge provides us with solutions to the Schrödinger equation that are separable in $x$ and $y$ coordinates, i.e., we may express $\psi(x,y)$ as $\psi(x,y)=\phi(x)\chi(y)$ if $\psi(x,y)$ is a viable solution. 

Let us suppose we have a reservoir, which is rotated by an arbitrary angle $\theta$ and centered at $(x_0,y_0)$ with respect to the origin, which we assume to be the center of the central region. We are then provided with the reservoir eigenfunctions of the form $\tilde{\psi}(\tilde{x},\tilde{y})$, where
\begin{equation}
\begin{aligned}
\tilde{x}(x,y) &= (x-x_0)\cos{\theta} + (y-y_0)\sin{\theta} \\
\tilde{y}(x,y) &= -(x-x_0)\sin{\theta} + (y-y_0)\cos{\theta}.
\end{aligned}
\end{equation}

\begin{figure}[ht]
    \centering
    \includegraphics[scale=1.4]{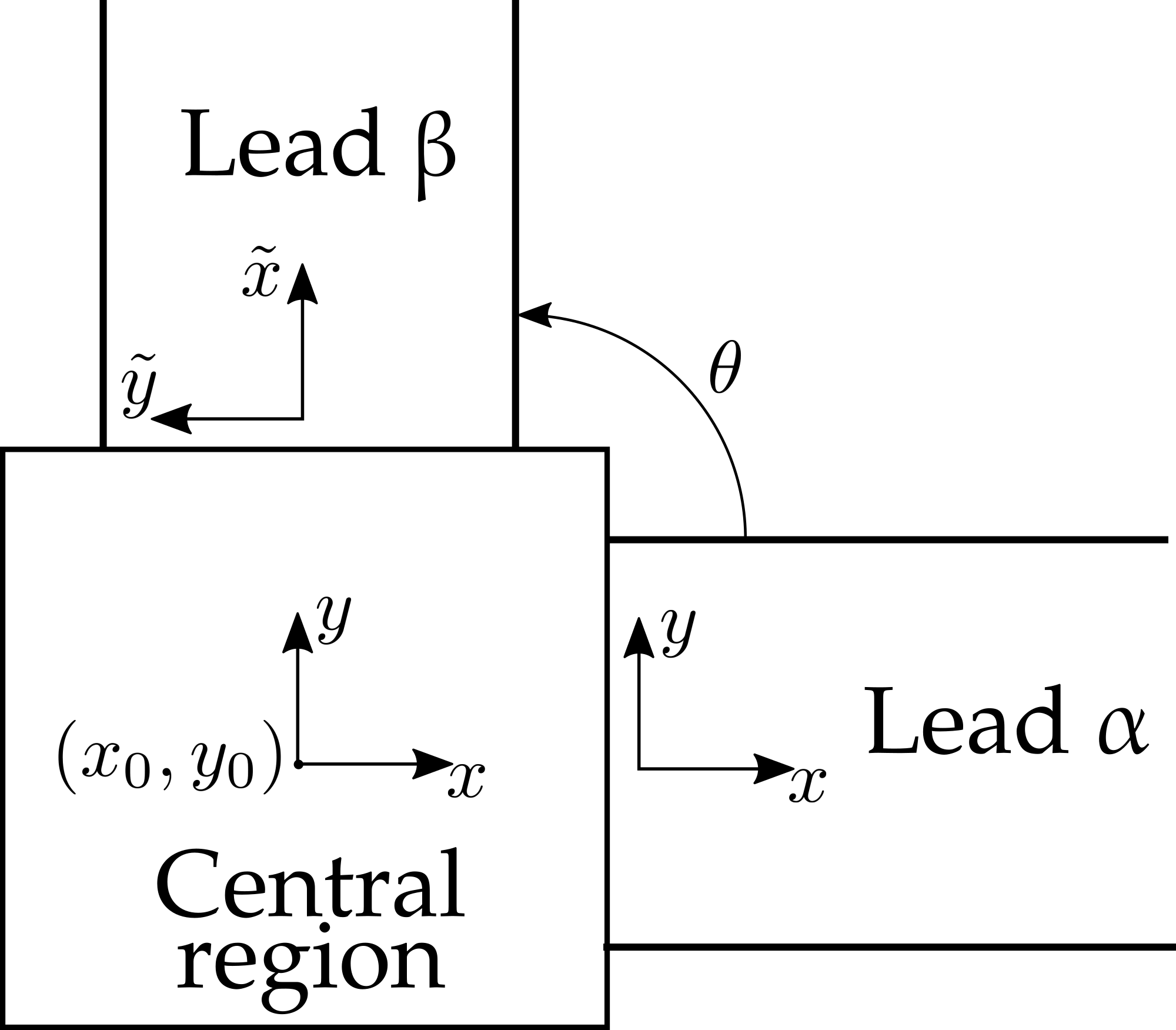}
    \caption{Schematic representation of a multi-lead system and the coordinate bases within the two leads. Angle $\theta$ represents the angle between the two bases, and is in this case $90^{\circ}$, but may be arbitrary in general.}
    \label{fig:gauge}
\end{figure}

Our task is to map the eigenfunction $\tilde{\psi}(\tilde{x},\tilde{y})$ with the vector potential $\tilde{\mathbf{A}}=-B\tilde{y}\hat{\tilde{x}}$ into an eigenfunction $\psi(x,y)$ with the vector potential $\mathbf{A}=-By\hat{x}$, same as for the central region. Figure \ref{fig:gauge} shows an example of how the coordinate systems in the leads are related. Since we already know how to map the coordinates from one frame to another, we only need to take care of the gauge transformation of the vector potential. Such a transformation is represented by a complex phase shift $e^{\mathrm{i}\Lambda(x,y)}$ of the eigenfunction. Here, we can write~\cite{Baranger1989} that
\begin{equation}
\begin{split}
\Lambda(x,y) = -&B(x-x_0)(y-y_0) \sin^{2}(\theta) \\ & + \frac{1}{4}B((y-y_0)^{2} - (x-x_0)^{2})\sin(2\theta).
\end{split}
\end{equation}

To summarize, our gauge transformation of the vector potential has a form
\begin{equation}\label{gauge_transform}
\psi(x,y) = e^{\mathrm{i}\Lambda(x,y)} \tilde{\psi}(\tilde{x}(x,y),\tilde{y}(x,y)).
\end{equation}
We want to emphasize that an arbitrary lead configuration can be taken into account by utilizing this gauge transformation~\cite{Baranger1989} when the transport system and the leads are affected by the magnetic field.

The above case of a perpendicular, uniform and constant magnetic field is a common setup in quantum transport
experiments (see, e.g., Refs.~\cite{Datta1997, reimann_rev.mod.phys_74_1283_2002, kouwenhoven_rep.prog.phys_64_701_2001}). 
If the magnetic field is non-zero in the leads, we choose the vector potential according to the linear gauge for quantum devices with multiple leads arranged in an arbitrary fashion, as it has been shown in Ref.~\cite{Baranger1989}. In the case of a {\em tilted} magnetic field this is not possible as we cannot utilize the gauge transformation in Eq.~(\ref{gauge_transform}). A typical numerical trick is to investigate a hybrid system composed of the original central region and the portion of the attached leads, where the magnetic field is present, or is changing, e.g., decaying eventually to zero. Then, we study the transport through the hybrid system coupled to the portion of the leads, where the magnetic field is absent. However, this approach requires the use of an eigenvalue solver for the central region that can handle a tilted magnetic field, or a magnetic
field varying in space. In principle, \textsc{tinie}'s software design allows this approach, although the present
approach demonstrated below resorts to the use of the \textsc{itp2d} software~\cite{Itp2d}, which is
restricted to 2D problems for perpendicular and homogeneous magnetic fields. As stated below, this approach
covers the majority of, e.g., magnetoconductance experiments.

\section{Design of the program}
\label{sec:implementation}

As mentioned in the introduction, \textsc{tinie} is designed for performance and interoperability, as well as for flexibility and ease-of-use. These features are combined by employing the features of the Python programming language as well as optimized algorithms and simple data structures. 

In \textsc{tinie}, the computation can be viewed a two-step process. First, in the system initialization phase, we determine the central region, i.e, the quantum device, the leads attached to the device, and the coupling between the central region and the leads. This phase is described in Secs.~\ref{sec:centralregion} and~\ref{ssec:coupling}. In the second step, the transport properties, such as the transmission or local density of states, are computed as described in Sec.~\ref{sub:quantumtransport}. This division enables more flexibility alongside the modular structure to extend \textsc{tinie} in the future, e.g., additions of other coupling schemes or eigenvalue solvers. The implementation of \textsc{tinie} is discussed in Sec.~\ref{sub:programstructure}--\ref{sub:comparison}.

\subsection{Central region}

\label{sec:centralregion}
\subsubsection{Discretization of eigenfunctions, potentials and coupling}
\label{ssec:discretizationeig}
Due to the nature of numerical computations, we discretize all the eigenfunctions of the QD and reservoirs by evaluating them on a 2D grid, transforming $\psi(x, y)$ into $\mathbf{\Psi}$:
$$
\mathbf{\Psi} = 
\left[\begin{array}{cccc}
 \psi(x_0, y_{N-1}) & \psi(x_1, y_{N-1})     & \cdots & \psi(x_{M-1},y_{N-1})      \\
 \psi(x_0, y_{N-2})     & \psi(x_1, y_{N-2}) & \cdots & \psi(x_{M-1},y_{N-2}) \\
 \vdots & \vdots & \ddots & \vdots \\
 \psi(x_0, y_0)     &  \psi(x_1, y_0)  & \cdots & \psi(x_{M-1},y_0) 
\end{array}\right] ,
$$
where $x_0, x_1, \ldots , x_{M-1}$ are the values of $x$-axis, discretized over $M$ equally-spaced points and $y_0, y_1, \ldots , y_{M-1}$ are the values of $y$-axis, discretized over $N$ equally-spaced points. Below we denote discretized eigenfunctions of any region as $\mathbf \Psi$. This procedure provides us with a set of discretized QD eigenfunctions $\{ \mathbf{\Psi}_C \}$ and a set of discretized reservoir eigenfunctions for each reservoir in the system $\{ \mathbf{\Psi}_{L_{\alpha}} \}$.

In the same manner we discretize the potential $V_{\mathrm{pot}}(x,y)$, so that it becomes $\mathbf{V}_{\mathrm{pot}}$:
$$
\mathbf{V}_{\mathrm{pot}} = 
\left[\begin{array}{cccc}
 V_{\mathrm{pot}}(x_0, y_{N-1}) & V_{\mathrm{pot}}(x_1, y_{N-1})     & \cdots & V_{\mathrm{pot}}(x_{M-1},y_{N-1})      \\
 V_{\mathrm{pot}}(x_0, y_{N-2})     & V_{\mathrm{pot}}(x_1, y_{N-2}) & \cdots & V_{\mathrm{pot}}(x_{M-1},y_{N-2}) \\
 \vdots & \vdots & \ddots & \vdots \\
 V_{\mathrm{pot}}(x_0, y_0)     &  V_{\mathrm{pot}}(x_1, y_0)  & \cdots  & V_{\mathrm{pot}}(x_{M-1},y_0) 
\end{array}\right].
$$

Next, we discretize the calculation procedure for the coupling matrix. Here we use the following approximation of Eq. (\ref{eq:coupling}):
\begin{equation}
\begin{split}
  \left[\mathbf{V}_{\alpha}\right]_{ij} \approx \int_{\Omega} -\frac{1}{2}&\mathbf{\Psi}_{L_{\alpha},i}^* \circ \left[\nabla^2 \mathbf{\Psi}_{C,j} + \frac{1}{2}\mathrm{i}By\frac{\partial}{\partial x} \mathbf{\Psi}_{C,j} \right] \, \\ & + \mathbf{\Psi}_{L_{\alpha},i}^* \circ \left[\mathbf{V}_{\mathrm{pot}} + \frac{1}{2}B^{2}y^{2}\right] \circ \mathbf{\Psi}_{C,j} \, \mathrm{d} \mathbf{r},
\end{split}
\end{equation}
where $\mathbf{\Psi}_{L_{\alpha},i}$ is the $i$th eigenfunction of the reservoir $\alpha$, $\mathbf{\Psi}_{C,j}$ is the $j$th eigenfunction of the central region, $\mathbf{V}_{\mathrm{pot}}$ is the disretized potential energy function of the overlapping region, $\circ$ is the Hadamard element-wise matrix product operator, and $\Omega$ is the region of overlap between the central region and the reservoir. We use the finite-difference methods implemented in the Python package \texttt{findiff}~\cite{FINDIFF}
to numerically evaluate the Laplacian $\nabla^{2}$ and the partial derivative $\frac{\partial}{\partial x}$. The numerical integration is performed with \texttt{scipy}'s Simpson's rule integration routines \cite{Scipy}.

\subsubsection{Discretization of Hamiltonians}
\label{ssec:discretizationham}
We select our eigenfunction sets in such a way that they form an orthonormal basis for their respective eigenspaces. Hence, the Hamiltonians for both the central region and the reservoir are diagonal. We then define our Hamiltonian operators in a matrix form for the central region:
$$
\mathbf{H}_C = \mathrm{diag}(E^C_0, E^C_1, \ldots, E^C_{N_{\mathrm{ctr}}-1}),
$$
where $\mathbf{H}_C \in \mathbb R^{N_\mathrm{ctr}\times N_\mathrm{ctr}}$, $N_\mathrm{ctr}$ is the total number of states in the central region, and $\{E^{C}_{j}\}_{j=0}^{N_\mathrm{ctr}-1}$ are the corresponding eigenenergies of the central region. Similarly for the reservoirs we have \\
$$
\mathbf{H}_{L_{\alpha}} = \mathrm{diag}(E^{L_{\alpha}}_0, E^{L_{\alpha}}_1, \ldots, E^{L_{\alpha}}_{N_{res,\alpha}-1}),
$$
where $\mathbf{H}_{L_{\alpha}} \in \mathbb R^{N_{\mathrm{res}, \alpha}\times N_{\mathrm{res}, \alpha}}$, $N_{\mathrm{res}, \alpha}$ is the total number of states in the reservoir $\alpha$, and $\{E^{L_{\alpha}}_{i}\}_{i=0}^{N_{\mathrm{res}, \alpha}-1}$ are the corresponding eigenenergies of the reservoir $\alpha$. This procedure provides us with a matrix form Hamiltonian of the QD and each of the reservoirs.

\subsection{Coupling matrix}
\label{ssec:coupling}

While the Hamiltonians and eigenfunctions of the QD and reservoirs are calculated in a straightforward
fashion based on the definition of the system, the calculation of the coupling matrix -- necessary
for transport calculations described in Sec. \ref{ssec:eset} -- requires additional elaboration. 
There are several ways to calculate the coupling. Many transport codes resort to the tight-binding coupling model \cite{Slater1954}, which is an approximation to the analytical form of the coupling. 
The coupling matrix elements can be written in an exact form as
\begin{equation}
\label{eq:couplinggeneral}
\left[\mathbf{V}_{\alpha}\right]_{ij} = \langle \psi_{L_{\alpha},i} \vert \hat{H} \vert \psi_{C,j} \rangle,
\end{equation}
where $\psi_{L_{\alpha},i}$ is the $i$th eigenfunction of the reservoir that we couple to the central region and $\psi_{C,j}$ is the $j$th eigenfunction of the central region. The Hamiltonian operator $\hat{H}$ for a particle of charge $q$ in a magnetic field has a general form~\cite{Goldstein2002} of 
\begin{equation}
    \hat{H} = \frac{1}{2}(\hat{p} - q\hat{A})^2 + \hat{V}_{\mathrm{pot}},
\end{equation}
where $\hat{p}=-\mathrm{i}\nabla$ is the momentum operator, $\hat{A}$ is the vector potential operator, and $\hat{V}_{\mathrm{pot}}$ is the potential operator. 

\subsubsection{Overlap coupling}
We may consider the coupling caused by the overlap of wave functions in the the region between the reservoir and the central region. In this case, with the choice of gauge defined in Sec.~\ref{ssec:bfield}, the coupling of Eq.~(\ref{eq:couplinggeneral}) has the form
\begin{equation}
\begin{split}
\label{eq:coupling}
    \left[\mathbf{V}_{\alpha}\right]_{ij} = \int\limits_{\Omega} \psi^{*}_{L_{\alpha},i}(\mathbf r) \bigg[-&\frac{1}{2}\nabla^2 + \frac{1}{2}\mathrm{i} B y \frac{\partial}{\partial x} \\ & + \frac{1}{2} B^{2} y^{2} + V_{\mathrm{pot}}(\mathbf r)\bigg] \psi_{C,j}(\mathbf r) \, \mathrm{d}\mathbf{r},
\end{split}
\end{equation}
where $\Omega$ is the region where our descriptions of the reservoir $\alpha$ and the central region overlap. In case the system contains more than one reservoir that is coupled to the central region, we need to calculate the coupling matrices for all the reservoirs in the system.

\subsubsection{Tight-binding coupling}
In some physical scenarios, it may also be plausible to look at the coupling caused by the tunneling of electrons from the reservoirs to the center region and the other way around. Such a coupling is incorporated into the tight-binding model~\cite{Datta1997, Stefanucci2013} as follows~\bibnote{This coupling scheme should not be confuse to originate from a nonlocal potential. Instead, it stems from utilizing tight-binding approximation, where we couple points between the central region and the corresponding lead in a weakly couple manner. However, we formulate this coupling scheme in the continuum limit, and it can be interpret as a generalization of the hopping elements in the nearest neighbour-coupling applied for a discrete real space representation. The effect of a magnetic field is stated very simply: the coupling element connecting two points $\mathbf{r}$ and $\mathbf{r}'$ is modified by a phase factor proportional to the line integral of the employed vector potential along the path connecting the points.}:
\begin{equation}
    \left[\mathbf{V}_{\alpha}\right]_{ij} = \int\limits_{\Omega_{L_{\alpha}}} \int\limits_{\Omega_{C}} \frac{\psi^{*}_{L_{\alpha},i}(\mathbf r')\psi_{C,j}(\mathbf r)}{\left\lVert \mathbf{r'} - \mathbf{r} \right\rVert^{2}} e^{-\mathrm{i}\theta} \, \mathrm{d}\mathbf{r} \, \mathrm{d}\mathbf{r'},
\end{equation}
where $\Omega_{C}$ is the section of the central region that we couple to the section of the reservoir $\alpha$, $\Omega_{L_{\alpha}}$. The phase factor $e^{-\mathrm{i}\theta}$ comes from the Peierls substitution that accounts for the inclusion of the vector potential in the system (see, e.g., Refs.~\cite{Peierls_I, Peierls_II, Peierls_III}). In our case, the magnetic field is constant and perpendicular to the plane, leading to a particularly simple form of the factor $\theta$ with the choice of gauge described in Sec. \ref{ssec:bfield}:
\begin{equation}
    \theta = -\frac{B}{2}(x' - x)(y' - y).
\end{equation}
\subsection{Quantum transport calculation}
\label{sub:quantumtransport}

\subsubsection{Self-energy calculator}
From Eq.~(\ref{eq:transmission}) we deduce that in order to obtain the transmission and currents we need the rate operators of the leads ($\hat{\Gamma}_{\alpha}(\omega)$) and the retarded/advanced Green's functions ($\hat{G}^{R/A}(\omega)$). Equations (\ref{eq:green}) and (\ref{eq:rateoperator}) suggest that we first need to compute the advanced and retarded self-energies. Using Eq.~(\ref{eq:selfenergy}) we obtain
\begin{equation}\label{eq:selfenergyterm}
\mathbf{\Sigma}^{R/A}_{\alpha}(\omega) =\mathbf{V}_{\alpha}^{\dagger}\left[(\omega \pm \mathrm{i}\eta)\mathbb{1}-\mathbf{H}_{{L}_{\alpha}}\right]^{-1}\mathbf{V}_{\alpha},
\end{equation}
where $\mathbb{1}$ is an identity matrix. The probe energy $\omega$ is discretized according to the user-specified energy spacing $\mathrm{d} \omega$. Because we represent the Hamiltonian operators in the eigenbasis, the matrices $(\omega \, \pm \, \mathrm{i}\eta)\mathbb{1} - \mathbf{H}_{{L}_{\alpha}}$ are diagonal, making them computationally easy to invert. With $\mathbf{\Sigma}^{R/A}_{\alpha}(\omega)$ known, we get the rate operators
\begin{equation}
\mathbf{\Gamma}_{\alpha}(\omega) = \mathrm{i}\left[\mathbf{\Sigma}^{R}_{\alpha}(\omega) - \mathbf{\Sigma}^{A}_{\alpha}(\omega)\right]
\end{equation}

In the calculation of the self-energies, we have defined in Eq.~(\ref{eq:selfenergyterm}) a positive $\eta$-parameter to incorporate the boundary conditions, and thus to distinguish the advanced and retarded Green functions. Theoretically, the parameter should be infinitesimal and eventually we should go to limit where $\eta$ vanishes. However, we require a finite $\eta$-value for numerical stability in order to prevent a divergence in the neighborhood of the eigenenergies of the Hamiltonian. This causes the shifting and broadening of energy levels, which can lead to missing out some essential features of the model. On the other hand, setting the parameter to very small values can cause numerical problems near the eigenenergies. In practice, the 
effect of finite $\eta$ should be regulated numerically.

The self-energies defined in Eq.~(\ref{eq:selfenergyterm}) describe the effect of the lead onto the quantum device. Instead of using Eq.~(\ref{eq:selfenergyterm}) directly, the self-energies can be calculated either with recursive methods~\cite{wenzien_j_phys_condens_matter_49_9893_1989, kudrnovsk_phys.rev.b_46_4222_1992, nardelli_phys.rev.b_60_7828_1999} or by utilizing semianalytic formulae~\cite{taylor_phys.rev.b_63_245407_2001, khomyakov_phys.rev.b_72_035450_2005, sanvito_phys.rev.b_59_11936_1999, umerski_phys.rev.b_55_5266_1997, ando_phys.rev.b_44_8017_1991,krstic_phys.rev.b_66_205319_2002}. The recursive methods suffer from a poor convergence rate in some transport systems, for example if the lead Hamiltonians are relatively sparse. On the other hand, the semianalytical approaches overcome the problem of the recursive methods by their construction. Nevertheless, major issues can arise when the coupling (hopping) matrices are singular or otherwise irregular. However, there is a robust semianalytical algorithm~\cite{rungger_phys.rev.b_78_035407_2008} overcoming the limitations which can be employed in conjunction with ab initio transport quantum transport calculations. In particular, this approach eliminates the need for a global finite $\eta$ parameter, such as that in Eq.~(\ref{eq:selfenergyterm}). Thanks to the flexible structure, these different methods can be employed with \textsc{tinie}.

We point out that \textsc{tinie} enables an optional user-defined coupling, i.e., the user specifies either the self-energies or the rate operators for each lead. This allows the user to bypass the actual coupling element calculations, which can be computationally demanding. Here the user does not need to specify the $\eta$-parameter. The coupling parameters may be obtained from experimental data and numerical calculations such as the algorithm described in Ref.~\cite{rungger_phys.rev.b_78_035407_2008}, or from analytic expressions, such as in the case of the wide-band limit approximation~\cite{Stefanucci2013} or semi-infinite leads~\cite{Datta1997}. In principle, this option allows \textsc{tinie} to compute quantum transport properties of quantum devices for arbitrary geometries and terminal configurations with irregularly shaped leads. Nonetheless, the self-energies or rate operators are, generally, not known, and thus \textsc{tinie} provides the functionality to determine the necessary coupling elements as described above.

\subsubsection{Green's functions calculator}
\label{sssec:greencalc}

With retarded self-energies calculated, we can now get the Green's functions. However, according to Eq.~(\ref{eq:green}) explicit numerical solution of the Green's function might be computationally expensive, as it requires an inversion of a dense matrix. A more efficient computational approach may be outlined as follows. First, we define the inverse retarded Green's function operator
\begin{equation}\label{eq:greenfunction}
[\mathbf{G}^{R}(\omega)]^{-1} = (\omega + \mathrm{i} \tilde{\eta}) \mathbb{1} - \mathbf{H}_C - \sum_{\alpha} \mathbf{\Sigma}^{R}_{\alpha}(\omega).
\end{equation}

We remind of the additional small term $\tilde{\eta}$, which is not present in Eq.~(\ref{eq:green}). We include this possibility as an additional measure to possible ensure the numerical stability. Again, the effect of a finite value should be minimized by setting $\tilde{\eta}$ close to zero. In contrast, the $\tilde{\eta}$ in Eq.~(\ref{eq:greenfunction}) can be significantly smaller than in the case of Eq.~(\ref{eq:selfenergyterm}) due to the self-energy term. In fact, the parameter $\tilde{\eta}$ is set to be zero as default. The actual need for $\tilde{\eta}$ depends on the self-energy, and hence on the type of the coupling. If this option is used, the effect of finite $\tilde{\eta}$ should be also regulated numerically.

The inverse advanced Green's function operator is defined similarly as $[\mathbf{G}^{A}(\omega)]^{-1} = ([\mathbf{G}^{R}(\omega)]^{-1})^{\dagger}$. Then, we observe that in Eq.~(\ref{eq:transmission}) we never explicitly need the Green's functions to compute the transmission, but rather we are only required to know their products with the rate operators. Hence, we use \texttt{scipy.linalg}'s \texttt{solve} routine to solve the linear equation $[\mathbf{G}^{R/A}(\omega)]^{-1} \mathbf{X} = \mathbf{\Gamma}_{\alpha}(\omega)$ for $\mathbf{X}$ efficiently. This gives us everything we need to calculate transmission and currents in the transport system.

\subsubsection{Transport properties calculator}

As the final step, we evaluate the transmission from reservoir $\alpha$ to reservoir $\beta$ by calculating $\mathbf{G}^{R}(\omega) \mathbf{\Gamma}_{\beta}(\omega)$ and $\mathbf{G}^{A}(\omega) \mathbf{\Gamma}_{\alpha}(\omega)$ as described in Sec. \ref{sssec:greencalc} and then using Eq.~(\ref{eq:transmission}). The calculation of $\mathcal{T}_{\alpha\beta}(\omega)$ for all possible reservoir pairs gives us the transmission matrix $\mathcal{T}$ evaluated at the probe energy $\omega$.

To compute the total current $I_{\alpha}$ running through each reservoir, we first compute the matrix of partial currents $i$. At this point, we extract the chemical potential $\mu$, transport system temperature $T$, and reservoir bias voltages $V_{\alpha}$ provided by the user to calculate the partial current matrix elements $i_{\alpha\beta}$ according to Eq.~(\ref{eq:pcurrents}). We use the \texttt{scipy}'s Simpson's rule routine for the integration. The integration boundaries are determined according to the chemical potential, reservoir bias voltages, and the broadening of the Fermi-Dirac distribution due to non-zero temperatures.

Additionally, after computing the transmission matrix we are able to evaluate the conductance matrix $\mathcal{G}$ using the result of Eq. (\ref{eq:conductance}). The transmission and the thermal broadening function are computed by using \texttt{numpy}'s routines.

\subsubsection{Density of states calculator}
In addition to the transport properties discussed in Sec. \ref{ssec:eset}, it is often beneficial to study how the energy states within the transport system are distributed. To that end, there are two quantities that provide us with crucial information: \textit{density of states} $g(\omega)$ (DOS) and \textit{local density of states} $\rho(\mathbf{r}, \omega)$ (LDOS). We may evaluate DOS directly from the retarded Green's function $\hat{G}^{R}(\omega)$ \cite{Datta1997}:
\begin{equation}
\label{eq:dos}
    g(\omega) = -\frac{1}{\pi}\mathrm{Tr}\left[\mathrm{Im}\left(\hat{G}^{R}(\omega)\right)\right].
\end{equation}

To compute the LDOS, we must first project the retarded Green's function of Eq. (\ref{eq:green}) into 
real space as
\begin{equation}
    \hat{G}^{R}(\mathbf{r}, \mathbf{r}, \omega) = \sum_{i,j} \psi_{C,i}^{*}(\mathbf{r}) \hat{G}^{R}(\omega) \psi_{C,j}(\mathbf{r}),
\end{equation}
where $\psi_{C,i}$ is the $i$th eigenfunction of the central region. The LDOS is then computed as \cite{Datta1997}:
\begin{equation}
\label{eq:ldos}
    \rho(\mathbf{r},\omega) = -\frac{1}{\pi} \mathrm{Im}\left[\hat{G}^{R}(\mathbf{r}, \mathbf{r}, \omega)\right].
\end{equation}

We conclude the description of \textsc{tinie}'s numerical routines with description of its procedure of DOS and LDOS computation. Unlike Sec. \ref{sssec:greencalc}, from Eq. (\ref{eq:dos}) we observe that in case of DOS/LDOS computations we need to know the value of the Green's function. As such, this time we can't go around computationally expensive inversion of a dense $[\mathbf{G}^{R}(\omega)]^{-1}$ matrix. Upon inverting the inverse of the retarded Green's function, we get retarded Green's function $\mathbf{G}^{R}(\omega)$. We may then use Eq. (\ref{eq:dos}) to obtain DOS $g(\omega)$:
\begin{equation}
    g(\omega) = -\frac{1}{\pi}\mathrm{Tr}\left[\mathrm{Im}\left(\mathbf{G}^{R}(\omega)\right)\right].
\end{equation}

To compute LDOS $\rho(\mathbf{r},\omega)$, we project $\mathbf{G}^{R}(\omega)$ into the real space numerically as follows:
\begin{equation}
    \mathbf{G}_{\mathbf{r}}^{R}(\omega) = \sum_{i,j} \left[\mathbf{G}^{R}(\omega)\right]_{ij} \left(\mathbf{\Psi}_{C,i}^{*} \circ \mathbf{\Psi}_{C,j}\right),
\end{equation}
then we use Eq. (\ref{eq:ldos}):
\begin{equation}
    \rho(\mathbf{r},\omega) = -\frac{1}{\pi}\mathrm{Im}\left[\mathbf{G}_{\mathbf{r}}^{R}(\omega)\right].
\end{equation}

\subsection{Program structure}
\label{sub:programstructure}

\begin{figure}[ht]
    \centering
    \begin{tikzpicture}[every node/.style={minimum height={1.5cm},thick,align=center,scale=0.9}]
    \node[draw, ellipse] (PT) {System \\ Initialization};
    \node[draw, below right=of PT] (CTR) {\texttt{Center} \\ object: \\ $\mathbf{H}_{C},\{\mathbf{\Psi}_{C}\}$};
    \node[draw, left=of CTR] (CPL) {\texttt{Coupling} \\ objects: \\ $\mathbf{V}_{\alpha}$};
    \node[draw, left=of CPL] (LD) {\texttt{Lead} \\ objects: \\ $\mathbf{H}_{L_{\alpha}},\{\mathbf{\Psi}_{L_{\alpha}}\}$};
    \node[draw, below=of CPL] (SYS) {\texttt{System} \\ object: \\ read/write};
    \node[draw, below=of SYS] (CALC) {\texttt{Calculator} \\ object: \\ compute};
    \node[draw, above right=of CALC] (SE) {\texttt{SelfEnergy} \\ object: \\ $\mathbf{\Sigma}^{R/A}_{\alpha},\mathbf{\Gamma}_{\alpha}$};
    \node[draw, below=of SE] (GF) {\texttt{GreenFunction} \\ object: \\ $\mathbf{G}^{R/A}$};
    \node[draw, ellipse, below right=of CALC] (RES) {Transport \\ Properties: \\ $\mathcal{T}_{\alpha\beta},\mathcal{G}_{\alpha\beta},i_{\alpha\beta},I_{\alpha}$};
    \node[draw, ellipse, below left=of CALC] (DOS) {DOS \\ and LDOS: \\ $g(\omega),\rho(\mathbf{r}, \omega)$};
    
    \draw[-{Latex[length=3mm]}] (PT) -- (CTR);
    \draw[-{Latex[length=3mm]}] (PT) -- (LD);
    \draw[-{Latex[length=3mm]}] (CTR) -- (CPL);
    \draw[-{Latex[length=3mm]}] (LD) -- (CPL);
    \draw[-{Latex[length=3mm]}] (CTR) -- (SYS);
    \draw[-{Latex[length=3mm]}] (CPL) -- (SYS);
    \draw[-{Latex[length=3mm]}] (LD) -- (SYS);
    \draw[-{Latex[length=3mm]}] (SYS) -- (CALC);
    \draw[-{Latex[length=3mm]}] (SE) -- (GF);
    \draw[-{Latex[length=3mm]}] (CALC) -- (SE);
    \draw[-{Latex[length=3mm]}] (GF) -- (CALC);
    \draw[-{Latex[length=3mm]}] (CALC) -- (RES);
    \draw[-{Latex[length=3mm]}] (CALC) -- (DOS);
    \end{tikzpicture}
    \caption{\textsc{tinie} object relation scheme.}
    \label{fig:algorithm}
\end{figure}
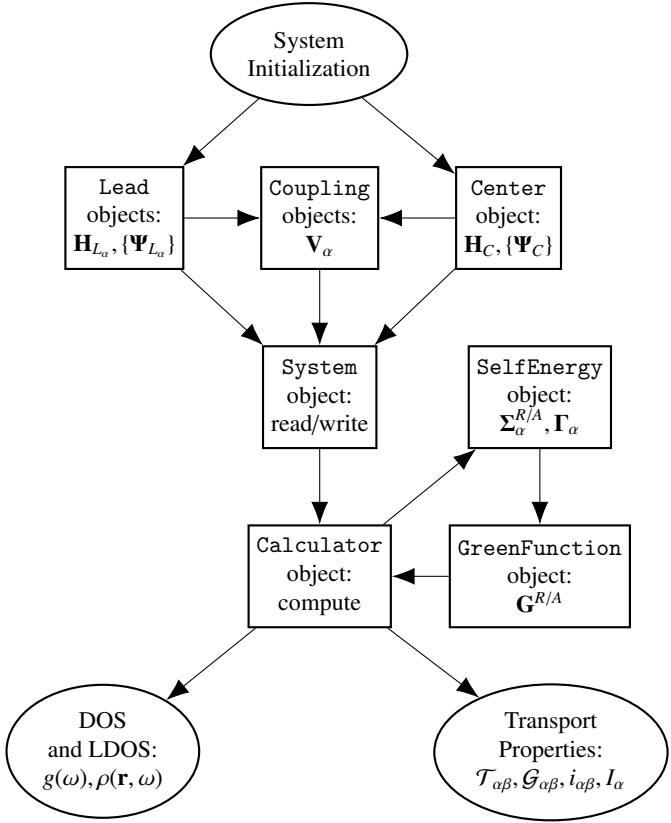

\textsc{tinie} has been written in Python 3.6, but the optimized numerical routines used in \texttt{numpy} and \texttt{scipy} have been inherited from C and C++. Thus, we have combined the readability of a high-level language such as Python with the high-performance computing features of a lower-level language such as C++. \textsc{tinie} is written in an object-oriented programming fashion, so all of the essential components of the code have been separated into different classes that interact with each other throughout the simulation process. The relations between the classes are summarized in Fig. \ref{fig:algorithm}.

The object types are listed in the following.
\begin{itemize}
    \item \texttt{Center} object: represents the central region in the transport system. It is responsible for computing/retrieving the central region Hamiltonian $\mathbf{H}_{C}$ and the set of eigenfunctions of the central region $\{\mathbf{\Psi}_{C}\}$. These eigenfunctions are represented by \texttt{numpy} 2D arrays.
    \item \texttt{Lead} object: represents a lead region in the transport system. It is responsible for computing/retrieving the lead region Hamiltonian diagonal matrix $\mathbf{H}_{L_{\alpha}}$, as well as its set of eigenfunctions $\{\mathbf{\Psi}_{L_{\alpha}}\}$.
    \item \texttt{Coupling} object: represents a coupling region between a lead and the central region. It is responsible for computing/retrieving the coupling $\mathbf{V}_{\alpha}$ between the central region and lead $\alpha$, which is represented by a \texttt{numpy} matrix array.
    \item \texttt{System} object: an interface that records the Hamiltonians and the coupling matrices in an HDF5 file via \texttt{h5py} module \cite{Collette2014}. As Hamiltonians and the coupling matrices do not change with the transport system parameters (such as the chemical potential or temperature), these files may then be reused for multiple transport calculations. Thus, \texttt{System} object also handles the retrieval of that data from an already existing file.
    \item \texttt{SelfEnergy} object: an interface that computes the self-energies $\mathbf{\Sigma}^{R/A}_{\alpha}$ and the rate operators $\mathbf{\Gamma}_{\alpha}$ for the transport system for varying values of the probe energy $\omega$.
    \item \texttt{GreenFunction} object: an interface that computes the Green's functions $\mathbf{G}^{R/A}$ for the transport system for varying the values of the probe energy $\omega$.
    \item \texttt{Calculator} object: an interface that performs the main transport calculation of partial currents $i_{\alpha\beta}$, total currents $I_{\alpha}$, transmission $\mathcal{T}_{\alpha\beta}$, and conductance $\mathcal{G}_{\alpha\beta}$.
\end{itemize}

All of the above-mentioned objects are \textit{abstract}, meaning that we can implement our own types of the central/lead/coupling region, specific to the transport system. This can be done by introducing a new class that inherits from one of those base classes and defining the respective methods for computation or retrieval of the system features.

Now we can outline the essential steps of the code execution:
\begin{enumerate}[Step 1:]
    \item System initialization step. Initializes the \texttt{Center} object and the \texttt{Lead} objects.
    \item Coupling step. \texttt{Coupling} objects are initialized and the coupling matrices between the regions are computed and stored in those objects.
    \item System finalization step. \texttt{Center}, \texttt{Lead} and \texttt{Coupling} objects are passed into the \texttt{System} interface to store the transport system data. This completes the setup of the transport system, preparing it for the subsequent transport calculations.
    \item Transport initialization step. \texttt{System} object is passed into the \texttt{Calculator} object to retrieve the transport system data.
    \item Calculator initialization step. Within the \texttt{Calculator}, \texttt{SelfEnergy} object is initialized. It is then passed into the \texttt{GreenFunction} object for its initialization.
    \item Transport calculation step. The system transmission, conductance, and currents are evaluated at the user-defined values for chemical potential, temperature, and lead biases using the self-energies, rate operators, and Green's functions.
\end{enumerate}

In essence, Steps 1-3 prepare the transport system for the calculation (\textsc{tinie\_prepare} stage), while the transport calculation itself is performed during Steps 4-6 (\textsc{tinie} stage). These two stages are thus completely independent from each other, as once the system is prepared and the system data stored in an HDF5 file, transport calculations with that file can be performed at will with varying chemical potential values, temperatures, or lead biases.

\subsection{Data files}

As described above, \textsc{tinie\_prepare} and \textsc{tinie} stages perform two independent parts of the transport calculation. Both of them produce their own data files so that the data can be processed at any time. We have chosen HDF5 (Hierarchical Data Format) for the data storage purposes. All the data of the simulations is saved in the HDF5 files. As the code has been written in Python, we have used the \texttt{h5py} package \cite{Collette2014} for the read/write HDF5 routines. Each HDF5 file produced by \textsc{tinie} stage has a 'type' attribute with the value \texttt{TINIEfile} and each HDF5 file produced by \textsc{tinie\_prepare} stage has a 'type' attribute with the value \texttt{PREPTINIEfile}. \texttt{PREPTINIEfile} HDF5 files store information such as the eigenfunctions, eigenenergies and the potentials of a prepared transport system. \texttt{TINIEfile} HDF5 files store information obtained from a transport calculation of the system, such as the transmission matrices, conductance, currents, and the calculation parameters, such as bias voltages, energy spacing, etc. More detailed information about the contents and the structure of the HDF5 files that \textsc{tinie} produces can be found on the \textsc{tinie}'s Gitlab project page. 

\subsection{Parallelization}

The coupling matrix elements can be computed independently of each other. This applies also to the transmission evaluated at the probe energy $\omega$. Hence, \textsc{tinie}'s routines have been parallelized by utilizing \texttt{mpi4py} \cite{Dalcin2005, Dalcin2008} -- Message Passing Interface for Python based on OpenMPI. When profiling the code, these two processes were also the ones that were the most time-consuming. The speed increase due to parallelization is roughly linearly proportional to the number of processors used for the computation.

\subsection{Comparison with other transport software}
\label{sub:comparison}

Quantum transport is addressed by several software packages in different domains. For example, there are quite a few packages, including commercial ones, for computing transport in molecular junctions. Examples include \textsc{transiesta}~\cite{Transiesta}, \textsc{smeagol}~\cite{Smeagol}, \textsc{openmx}~\cite{Openmx} and \textsc{nanodcal} along with \textsc{nanodsim}~\cite{nanodsim}. These packages combine density-functional theory with the non-equilibrium Green's function technique. Another category of transport codes is mainly geared towards the simulation of transistors on the nano- and mesoscale. This class contains packages such as \textsc{nemo5}~\cite{nemo5}, \textsc{nextnano}~\cite{Nextnano}, \textsc{nanotcad vides}~\cite{NanoTCAD}, and \textsc{tbsim}~\cite{TBsim}. An extension of these packages outside the scope of their specific class is often impossible or requires a lot of work. In addition to all these specialized packages, \textsc{kwant}~\cite{Kwant} offers a generic platform for a tight-binding quantum transport problem without being limited to a certain class of systems. All these packages go beyond the scope of \textsc{tinie} by considering more complicated physical effects such as involving phonon effects or self-consistent electrostatic potential calculations. 

Similarly to the principles of \textsc{kwant}, \textsc{tinie} emphasizes generality in order to compute the transport properties in various experimentally relevant two-dimensional nanostructures. With \textsc{tinie}, it is easy, for example, to study quantum dots with soft confining boundaries, which can be a difficult regime for quantum transport packages based on the tight-binding approach. In particular, \textsc{tinie} is designed for flexibility and ease-of-use as highlighted above. The modular design of the code enables for an easy expansion to include additional physical effects. Another advantage of \textsc{tinie} is the compatibility with external 2D software such as \textsc{itp2d}~\cite{Itp2d}. The open-source policy can drive the evolution of \textsc{tinie} further towards a more versatile quantum transport package.  

The current version of \textsc{tinie} allows us to investigate systems in homogeneous magnetic fields perpendicular to the transport setup. However, we acknowledge that some of the named quantum transport packages do allow to investigate systems in magnetic fields, even inhomogenous ones. For example, the generic transport software \textsc{kwant} relies on the tight-binding formulation and Peierl's substitution. Although this is a valid description in some nanostructures such as a molecular junction, \textsc{tinie} can go further by providing a more realistic scheme for a quantum device modeled by a smooth potential, e.g., a semiconductor quantum dot, in a (possibly very strong) magnetic field (see Ref.~\cite{Itp2d}). Similarly to transport software such as \textsc{kwant},\textsc{tinie} can also carry out quantum transport calculations even in the presence of a magnetic field in the leads. As discussed above, \textsc{tinie} can, in principle, handle tilted or inhomogeneous magnetic fields either by providing the necessary inputs, i.e, the Hamiltonians and the self-energies or equivalently rate operators, or by changing the eigenvalue solver to one that supports such functionality.

\section{Numerical testing}
\label{sec:tests}

\subsection{Automated testing framework}

We have implemented an automated testing framework for \textsc{tinie}. It probes all the basic functionalities of the package, makes sure that the functions of all the system classes behave as intended, and checks whether the asymptotic behavior of the numerical algorithms is correct. Most importantly, the framework compares the results of some simple transport calculations to their analytical solutions and checks the crucial symmetry properties of the transmission and partial current matrices.

\subsection{Test cases}
\label{ssec:testcases}

\subsubsection{One- and two-level systems}
\label{ssec:simplesys}
\begin{figure}[ht]
    \centering
    \includegraphics[scale=1.4]{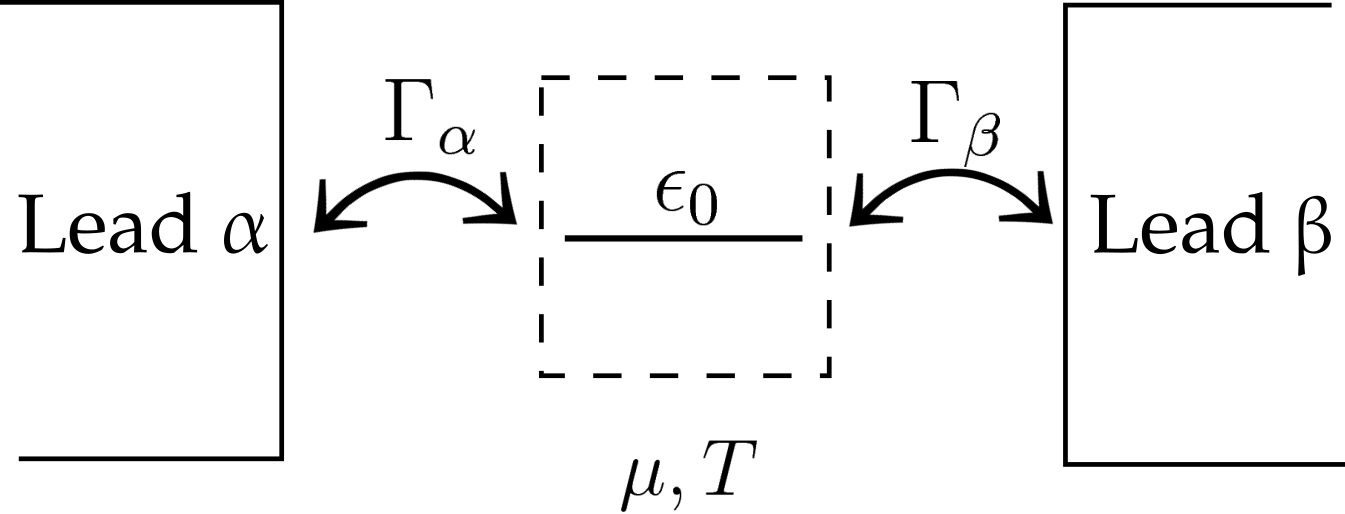}
    \caption{Schematic representation of a one-level transport system.}
    \label{fig:onestatefig}
\end{figure}
We start with one of the simplest possible transport systems: a single energy level connected to two leads. Figure \ref{fig:onestatefig} shows the structure of such a system. We utilize the \textit{wide band limit approximation} (WBLA) \cite{Stefanucci2013} to infer the transport properties. In a system that obeys WBLA, the rate operators are independent of the probe energy $\omega$. This allows us to bypass the computation of the coupling matrices. Instead, the coupling strength between the leads is then specified by the rate operators, which are constant with respect to $\omega$. \textsc{tinie} supports WBLA as it allows the user to specify custom rate operators when needed.

Overall, we need to know the following quantities to describe such a transport system: the energy of the center region $\epsilon_{0}$, rate operators $\Gamma_{\alpha}$ and $\Gamma_{\beta}$ for the leads $\alpha$ and $\beta$, respectively, bias voltages in the leads $V_{\alpha},V_{\beta}$, chemical potential $\mu$ and temperature $T$ of the system.
It can be shown that in the case of zero temperature and a small potential difference between the leads, we get the following analytical expressions for the transmission and current \cite{Ventra2008}:
\begin{equation}
\label{eq:onestate}
    \mathcal{T}_{\alpha\beta} = \frac{2 \pi}{2 \left[V_{\beta} - V_{\alpha}\right]} I_{\alpha} \quad \textrm{and} \quad I_{\alpha} = \frac{\Gamma_{\alpha} \Gamma_{\beta}}{(\mu - \epsilon_{0})^{2} + \frac{1}{4}(\Gamma_{\alpha} + \Gamma_{\beta})^2}.
\end{equation}
We benchmark the numerical precision of \textsc{tinie} in computing the current and transmission for different values of the lead rate operators, and compare the results to the analytical results at zero temperature. To test the performance of \textsc{tinie} at non-zero temperatures we thus compare the obtained results against exact (analytical) benchmark values.

We set $\epsilon_{0}=500$, $\mu=250$. Additionally, we set the bias potentials of the two leads to be $V_{\alpha}=0$ and $V_{\beta}=10^{-5}$. As the potential difference between the two leads is small, we compare our numerical results with the analytical result of Eq. (\ref{eq:onestate}). We set the lead rate operators $\Gamma_{\alpha}=\Gamma_{\beta}=\Gamma$ and compute the transmission and current in the system evaluated at various values of $\Gamma$. We start with the zero temperature case. Tables \ref{tbl:zerotrans} and \ref{tbl:zerocurrent} show the results with the relative error estimates.

\begin{table}[ht]
\centering
\renewcommand{\arraystretch}{1.25}
\caption{Comparison of the transmission values running through a zero-temperature one-state system computed using \textsc{tinie} against the analytical results of Eq. (\ref{eq:onestate}). Relative error tolerances are near zero due to limitations of finite-precision arithmetic.}
\label{tbl:zerotrans}
\begin{tabularx}{\linewidth}{c | >{\centering}X >{\centering}X  c}
\hline
$\Gamma$ & Analytical $\mathcal{T}_{\alpha\beta}(\mu)$ & \textsc{tinie} $\mathcal{T}_{\alpha\beta}(\mu)$ & Relative error\\
\hline \hline
$0.2$ & \mbox{$6.3999959\times10^{-7}$} & \mbox{$6.3999959\times10^{-7}$} & $\lesssim10^{-15}$ \\
$0.4$ & \mbox{$2.5599934\times10^{-6}$} & \mbox{$2.5599934\times10^{-6}$} & $\lesssim10^{-15}$ \\
$0.6$ & \mbox{$5.7599668\times10^{-6}$} & \mbox{$5.7599668\times10^{-6}$} & $\lesssim10^{-15}$ \\
$0.8$ & \mbox{$1.0239895\times10^{-5}$} & \mbox{$1.0239895\times10^{-5}$} & $\lesssim10^{-15}$ \\
$1.0$ & \mbox{$1.5999744\times10^{-5}$} & \mbox{$1.5999744\times10^{-5}$} & $\lesssim10^{-15}$ \\
\hline
\end{tabularx}
\end{table}

\begin{table}[ht]
\renewcommand{\arraystretch}{1.25}
\caption{Comparison of the current values for a zero-temperature one-state system computed using \textsc{tinie} against the analytical values of Eq. (\ref{eq:onestate}). The energy spacing for the numerical integration of Eq. (\ref{eq:pcurrents}) is set to $\mathrm{d}\omega=10^{-7}$.}
\label{tbl:zerocurrent}
\begin{tabularx}{\linewidth}{c | >{\centering}X  >{\centering}X  c}
\hline
$\Gamma$ & Analytical $I_{\alpha}$ & \textsc{tinie} $I_{\alpha}$ & Relative error\\
\hline \hline
$0.2$ & \mbox{$2.0371819\times10^{-12}$} & \mbox{$2.0371819\times10^{-12}$} & $1.910\times10^{-8}$ \\
$0.4$ & \mbox{$8.1487122\times10^{-12}$} & \mbox{$8.1487120\times10^{-12}$} & $1.910\times10^{-8}$ \\
$0.6$ & \mbox{$1.8334543\times10^{-11}$} & \mbox{$1.8334543\times10^{-11}$} & $1.910\times10^{-8}$ \\
$0.8$ & \mbox{$3.2594598\times10^{-11}$} & \mbox{$3.2594597\times10^{-11}$} & $1.910\times10^{-8}$ \\
$1.0$ & \mbox{$5.0928766\times10^{-11}$} & \mbox{$5.0928765\times10^{-11}$} & $1.910\times10^{-8}$ \\
\hline
\end{tabularx}
\end{table}

The values obtained numerically match the analytical values, especially in case of transmission. The relative error estimates are close to zero and limited by finite-precision arithmetics. For the current, the relative errors are very small as well. The minor deviations arise from the numerical integration over transmission values in the region $\left[V_{\alpha}, V_{\beta}\right]$. 

We note that both the transmission and the current increase with $\Gamma$. In the wide-band approximation regime, the value of $\Gamma$ corresponds to the strength of the coupling of the lead to the central region. Hence, our results are plausible: the stronger the coupling, the higher the transmission.

We have also investigated a non-zero temperature system with $T=100$. We have changed bias potentials to $V_{\alpha}=0$ and $V_{\beta}=100$. The other system parameters are the same as above, and once again we let $\Gamma$ vary. For such a system, there is no closed-form result that we can use for comparison. Instead, we have compares the results of \textsc{tinie} against numerically accurate benchmark results. We expect the current values to be higher in the non-zero temperature transport system due to the broader probe energy range and the thermal broadening of the Fermi-Dirac energy distribution. The results of the calculations are summarized below in Table \ref{tbl:nonzerocurrent}. We observe that the current values are much higher in this case of non-zero temperature, which supports our hypothesis.

\begin{table}[ht]
\renewcommand{\arraystretch}{1.25}
\caption{Comparison of the current values running through a non-zero temperature one-state system computed using \textsc{tinie} against the values computed numerically. The energy spacing for the numerical integration of the current over the probe energies has been set to $\mathrm{d}\omega=10^{-2}$.}
\label{tbl:nonzerocurrent}
\begin{tabularx}{\linewidth}{c | >{\centering}X  >{\centering}X  c}
\hline
$\Gamma$ & Analytical $I_{\alpha}$ & \textsc{tinie} $I_{\alpha}$ & Relative error\\
\hline \hline
$0.2$ & \mbox{$2.1314269\times10^{-2}$} & \mbox{$2.1314269\times10^{-2}$} & $2.824\times10^{-9}$ \\
$0.4$ & \mbox{$4.2630084\times10^{-2}$} & \mbox{$4.2630084\times10^{-2}$} & $5.749\times10^{-9}$ \\
$0.6$ & \mbox{$6.3947360\times10^{-2}$} & \mbox{$6.3947360\times10^{-2}$} & $8.702\times10^{-9}$ \\
$0.8$ & \mbox{$8.5266014\times10^{-2}$} & \mbox{$8.5266013\times10^{-2}$} & $1.160\times10^{-8}$ \\
$1.0$ & \mbox{$1.0658596\times10^{-1}$} & \mbox{$1.0658596\times10^{-1}$} & $1.484\times10^{-8}$ \\
\hline
\end{tabularx}
\end{table}

\begin{figure}[h]
    \centering
    \includegraphics[scale=1.4]{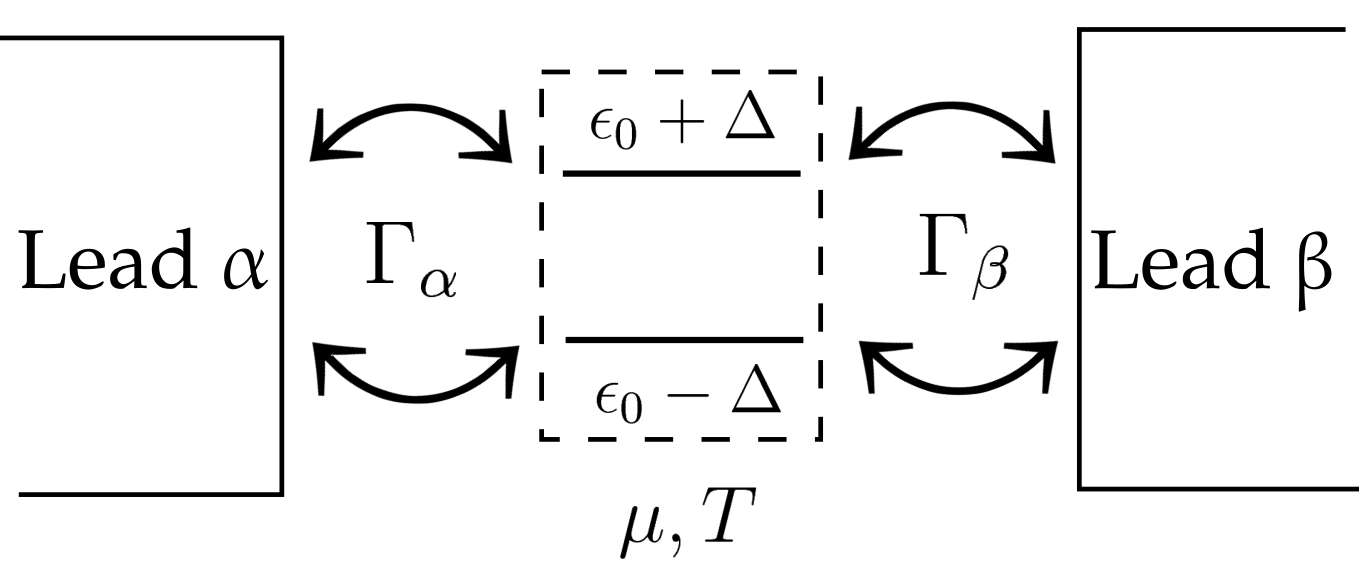}
    \caption{Schematic representation of a two-level transport system.}
    \label{fig:twostatefig}
\end{figure}

\begin{figure}[ht]
    \centering
    \includegraphics[scale=0.52]{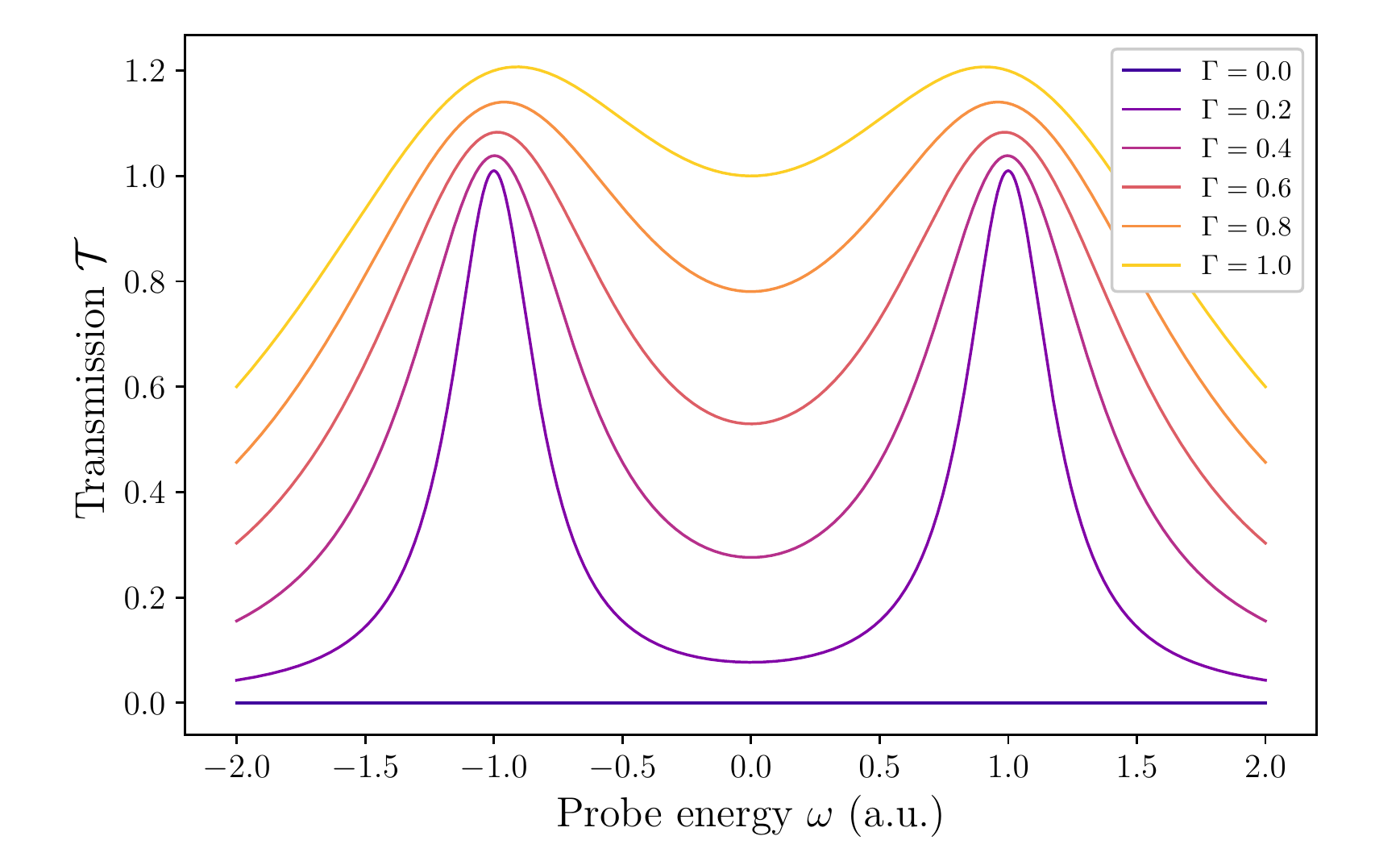}
    \caption{Transmission in a two-level transport system with varying strengths for the rate operators.}
    \label{fig:two_state}
\end{figure}

We now move on to a more realistic two-level molecular junction connected to two leads. Figure \ref{fig:twostatefig} shows the structure of such a system. We consider a center region that can be interpreted as a molecule, e.g., benzene, where the lower energy state is the highest occupied molecular orbital (HOMO), while the higher state is the lowest unoccupied molecular orbital (LUMO) \cite{Heurich2002}. We construct the central region Hamiltonian of the form
\begin{equation*}
H_{C} = 
\left[\begin{array}{cc} 
\epsilon_{0} + \Delta & 0 \\
0 & \epsilon_{0} - \Delta
\end{array}\right],
\end{equation*}
where $\epsilon_{0}$ is the Fermi energy and $\Delta$ is the parameter that we use to tune the energy spacing. Once again, we utilize WBLA, and the rate operators of the leads are presented as fixed matrices of the form \cite{Smeagol}
\begin{equation*}
\Gamma_{\alpha} = \Gamma_{\beta} = 
\left[\begin{array}{cc} 
\Gamma & 0 \\
0 & \Gamma
\end{array}\right],
\end{equation*}
where $\Gamma$ is the coupling strength parameter to be adjusted. 
To demonstrate how \textsc{tinie} handles such a two-level system, we compute its transmission, setting $\epsilon_{0}=0,\Delta=1,\mu=0$. Additionally, we set the lead bias voltages to $V_{\alpha}=-2, V_{\beta}=2$. We consider this system in a zero-temperature environment. We then investigate the behavior of the system as we vary $\Gamma$. We expect to observe peaks in transmission at the eigenenergies of the center ($\epsilon_{0}+\Delta=1$ and $\epsilon_{0}-\Delta=-1$), and we are interested in how those peaks vary with changing $\Gamma$. Figure \ref{fig:two_state} contains the results of the simulations.

As expected, we find peaks in the transmission at the eigenenergies of the central system. When $\Gamma=0$, the transmission is constantly zero, as in that case the leads are not coupled to the center at all. We can see that with increasing $\Gamma$ the peaks get broader and higher, and tend to merge together as $\Gamma$ goes to one. This result is plausible in view of the physical interpretation of $\Gamma$ considered above. By increasing $\Gamma$, we increase the energy bandwidth of the electrons that may pass through the central region, leading to the widening of the transmission peaks around the eigenenergies. Hence, \textsc{tinie} correctly captures the essential physical characteristics of the two-level system.

\subsubsection{Potential barrier}

\label{sec:barrier}
\begin{figure}[ht]
    \centering
    \includegraphics[scale=1.4]{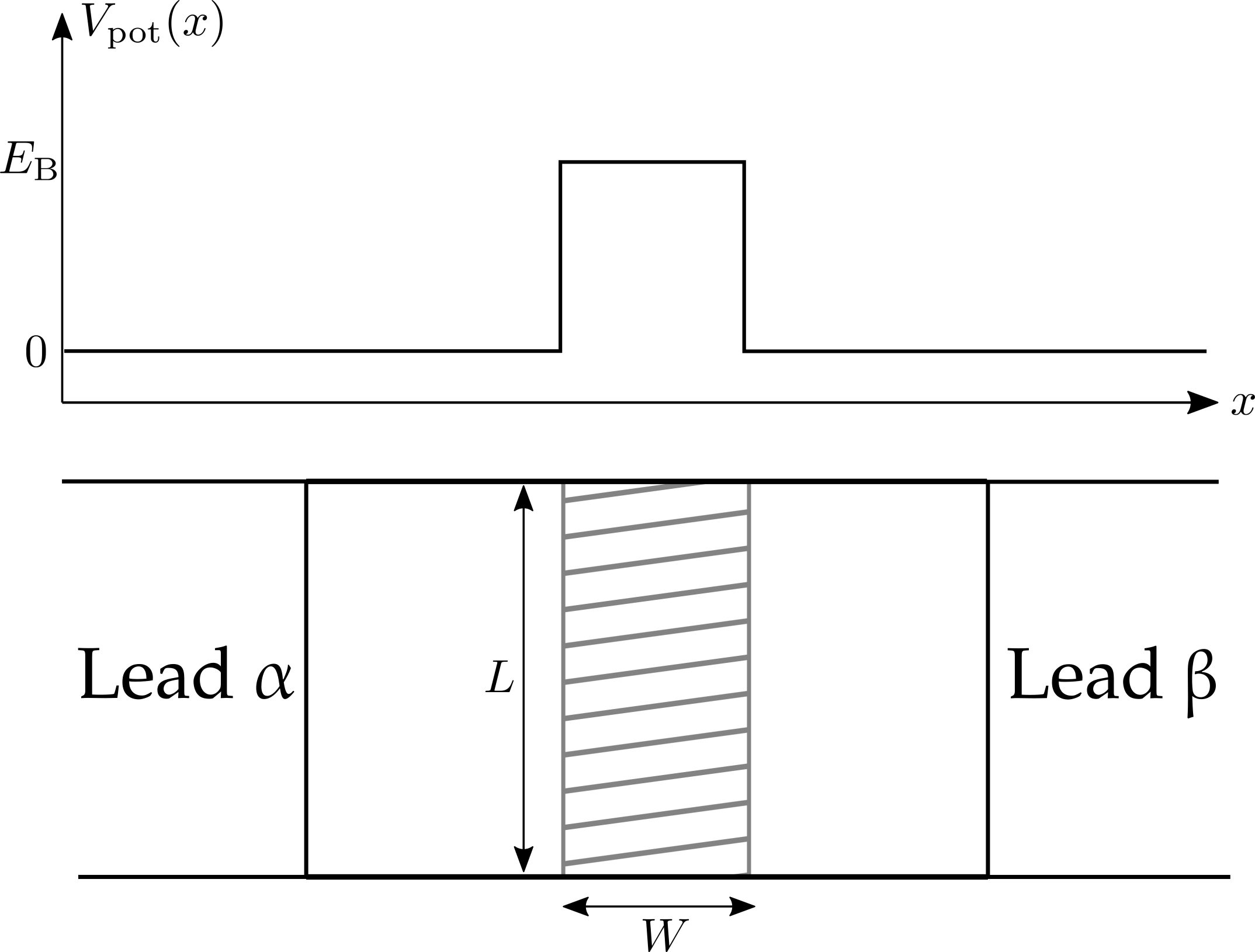}
    \caption{Schematic representation of the barrier potential in $x$-direction (top) and a two-dimensional potential barrier system (bottom).}
    \label{fig:barrierfig}
\end{figure}
Next we consider a conventional potential barrier system, which can be used to investigate electron tunneling properties across a nanostructure. The potential of the central region can be written as
\begin{equation}
\label{eq:barrier}
    V_{\mathrm{pot}}(x,y) = \begin{cases}
    E_{\mathrm{B}} & x \in [-\frac{W}{2},\frac{W}{2}] \wedge y \in [-\frac{L}{2},\frac{L}{2}] \\
    0 & x \in [-\frac{W}{2},\frac{W}{2}]^{C} \wedge y \in [-\frac{L}{2},\frac{L}{2}] \\
    \infty & \mathrm{elsewhere}.
    \end{cases}
\end{equation}
Here $E_{\mathrm{B}}$ is the barrier height, $L$ is the length of the central region in the $y$-direction and $W$ is the width of the barrier in the $x$-direction. Figure \ref{fig:barrierfig} illustrates $V_{\mathrm{pot}}$ with its key parameters. The eigenfunctions of the central region in this potential are solved using \textsc{itp2d} \cite{Luukko2013}. The central region is then connected to the system using the \texttt{Itp2dCenter} interface. 

Two leads are connected to the central region. The electrons are confined in the $y$-direction and propagating in the $x$-direction. We use a harmonic oscillator potential in the $y$-direction and a standard particle-in-a-box potential in the $x$-direction. The eigenfunctions for the leads can be solved analytically, leading to \cite{Datta1997}
\begin{equation}
    \psi_{k,l}^{L}(x,y) = \mathcal{N}\cos(k(x-x_{\mathrm{max}}^{L}) + \frac{\pi}{2}) e^{-\frac{1}{2} y^2} H_l(y),
\end{equation}
where $H_l(x)$ is the $l$th order Hermite polynomial and $\mathcal{N}$ is the normalization factor for the wave function. The indices $l$ and $k$ are the quantum numbers describing the system in $x$ and $y$ directions, respectively. The leads have been implemented in \textsc{tinie} as \texttt{FiniteHarmonicLead} object.

\begin{figure*}[ht]
    \centering
    \subfloat{\includegraphics[scale=0.52]{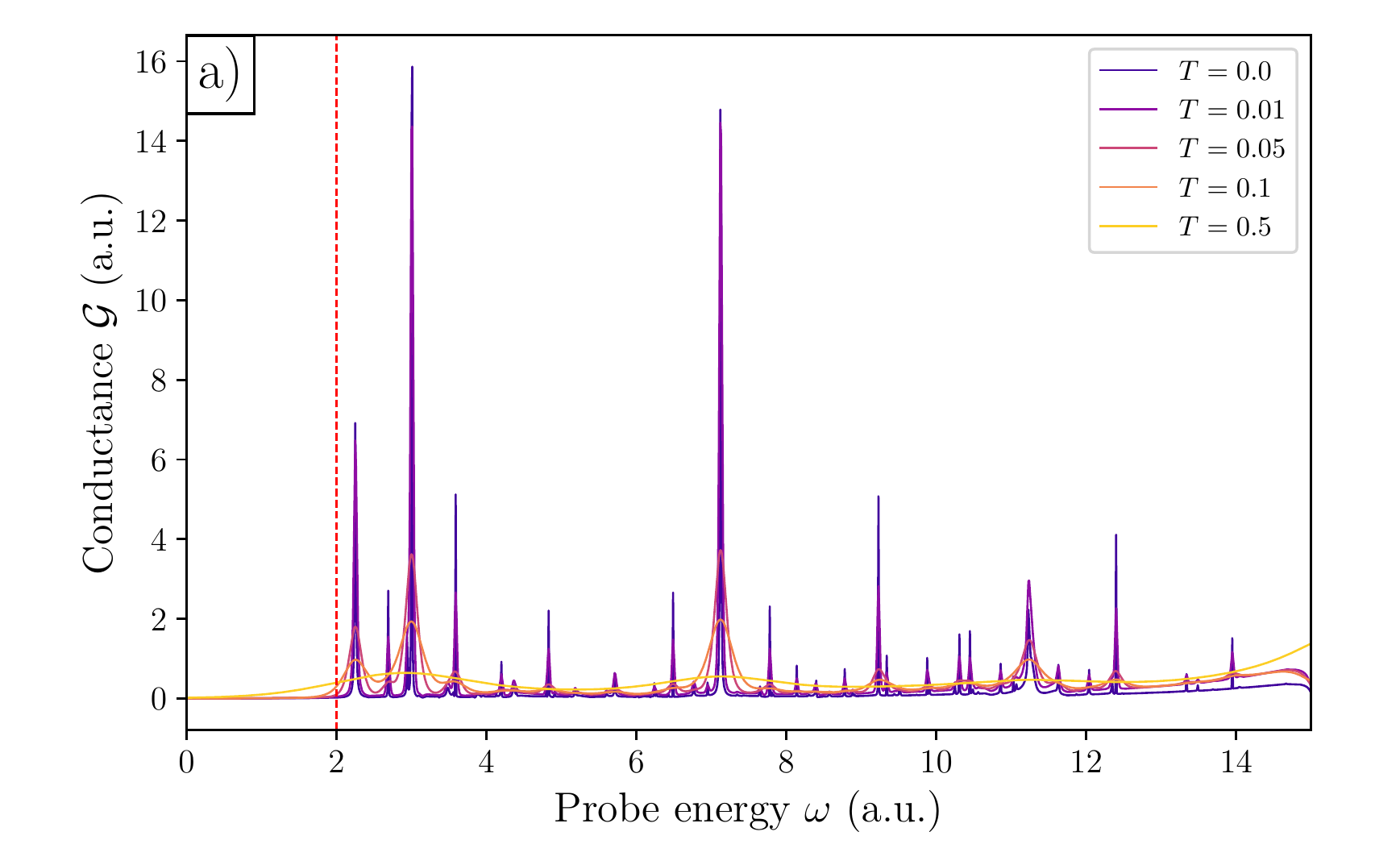}}
    \subfloat{\includegraphics[scale=0.52]{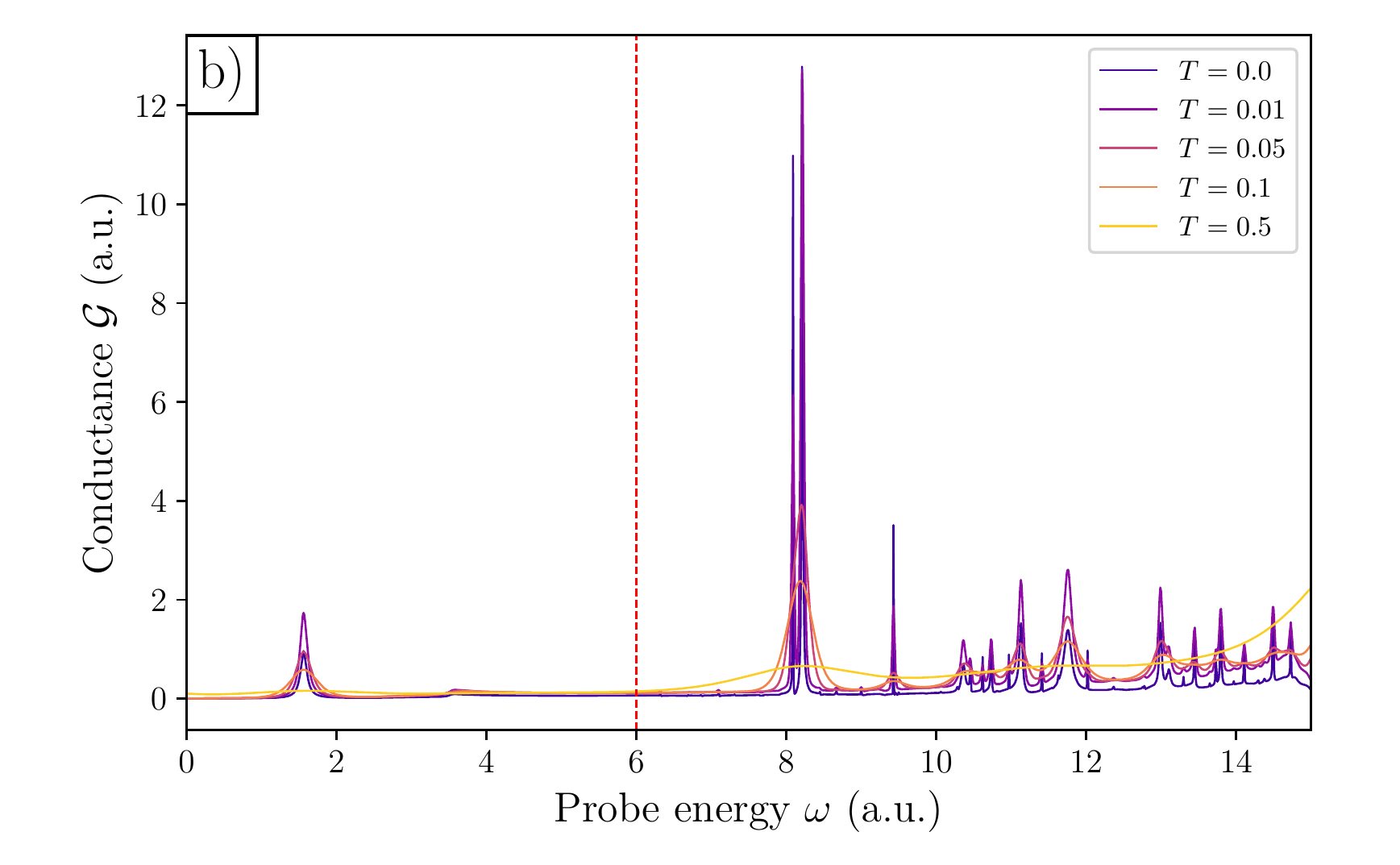}}
    \\
    \subfloat{\includegraphics[scale=0.52]{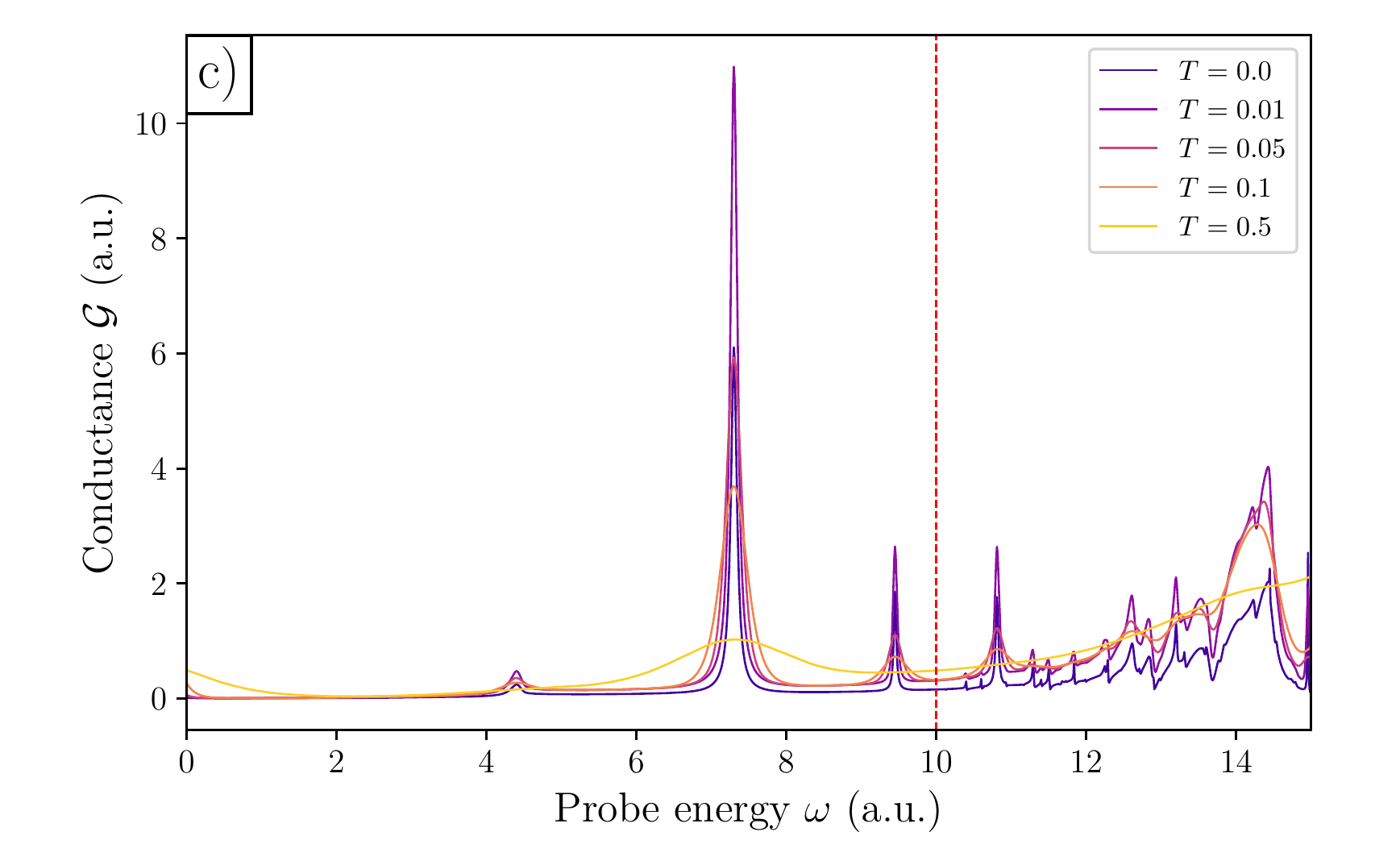}}
    \caption{Conductance in the potential barrier system at varying system temperatures with $\mu=0.0,\eta=0.02,V_{\alpha}=0.0,V_{\beta}=15.0$. The vertical red dotted line denotes the barrier energy $E_{\mathrm{B}}$, which has been set to 2, 6, and 10 in Figures a), b) and c) respectively. The conductance at zero temperature is scaled by a factor of 50 for visibility.}
    \label{fig:barrier}
\end{figure*}

The system has the following spatial confinements for the lead region and the center region:
\begin{itemize}
    \item Center region: $x,y \in [-6,6]$;
    \item Lead $\alpha$: $x \in [-100,0]$ and $y \in [-5,5]$;
    \item Overlap $\alpha$: $x \in [-6,0]$ and $y \in [-5,5]$;
    \item Lead $\beta$: $x \in [0,100]$ and $y \in [-5,5]$;
    \item Overlap $\beta$: $x \in [0,6]$ and $y \in [-5,5]$.
\end{itemize}
We set the width of the potential barrier to be 10, that is, $y \in [-5,5]$. We consider the behavior of conductance $\mathcal{G}$ from lead $\alpha$ to lead $\beta$, as we vary the barrier energy $E_{\mathrm{B}}$. Furthermore, we investigate the temperature effects on the conductance. Figure \ref{fig:barrier} shows the results of the simulations.

In these numerical studies, we have considered the probe energy range $\omega \in [0,15]$. In this energy range, each lead is found to contain 225$\,$000 eigenstates. As for the central region, solving its Schr{\"o}dinger equation with \textsc{itp2d} yields approximately 250 eigenstates in the same energy range. Consequently, this transport system is vastly more complex than the systems considered above in Sec. \ref{ssec:simplesys}.

We note that in Fig. \ref{fig:barrier}(a), the conductance only starts to grow and fluctuate when the energy of the probe electron surpasses that of the potential barrier. We observe similar behavior in Figs. \ref{fig:barrier}(b) and \ref{fig:barrier}(c). However, the picture is slightly more complex as demonstrated by the presence of minor conductance peaks below the barrier energy. They arise from the resonances in the system, as some of the eigenstates in the central region have energies below the potential barrier. The resonances occur when an eigenenergy of the lead closely matches one of the central region eigenenergies. 

When the temperature of the system is close to zero, or small relative to the Fermi energy, every resonance results in a Dirac delta function-like peak in the conductance. As the temperature is increased to the scales comparable with the Fermi energy, the peaks become broader and smaller due to the effects of the thermal broadening on the conductance [Eq. (\ref{eq:conductance})]. Thus, a single outlying conductance peak will get completely removed at high temperatures, while the peaks in the areas dense with conductance resonances will become more pronounced. This effect of temperature on conductance is observed in Fig. \ref{fig:barrier}, further reassuring us that \textsc{tinie} is capable of handling transport systems of high orders of complexity as well.

\subsubsection{Two-dimensional potential well in a magnetic field}
\label{sec:fockdarwin}
\begin{figure}[ht]
    \centering
    \includegraphics[scale=0.075]{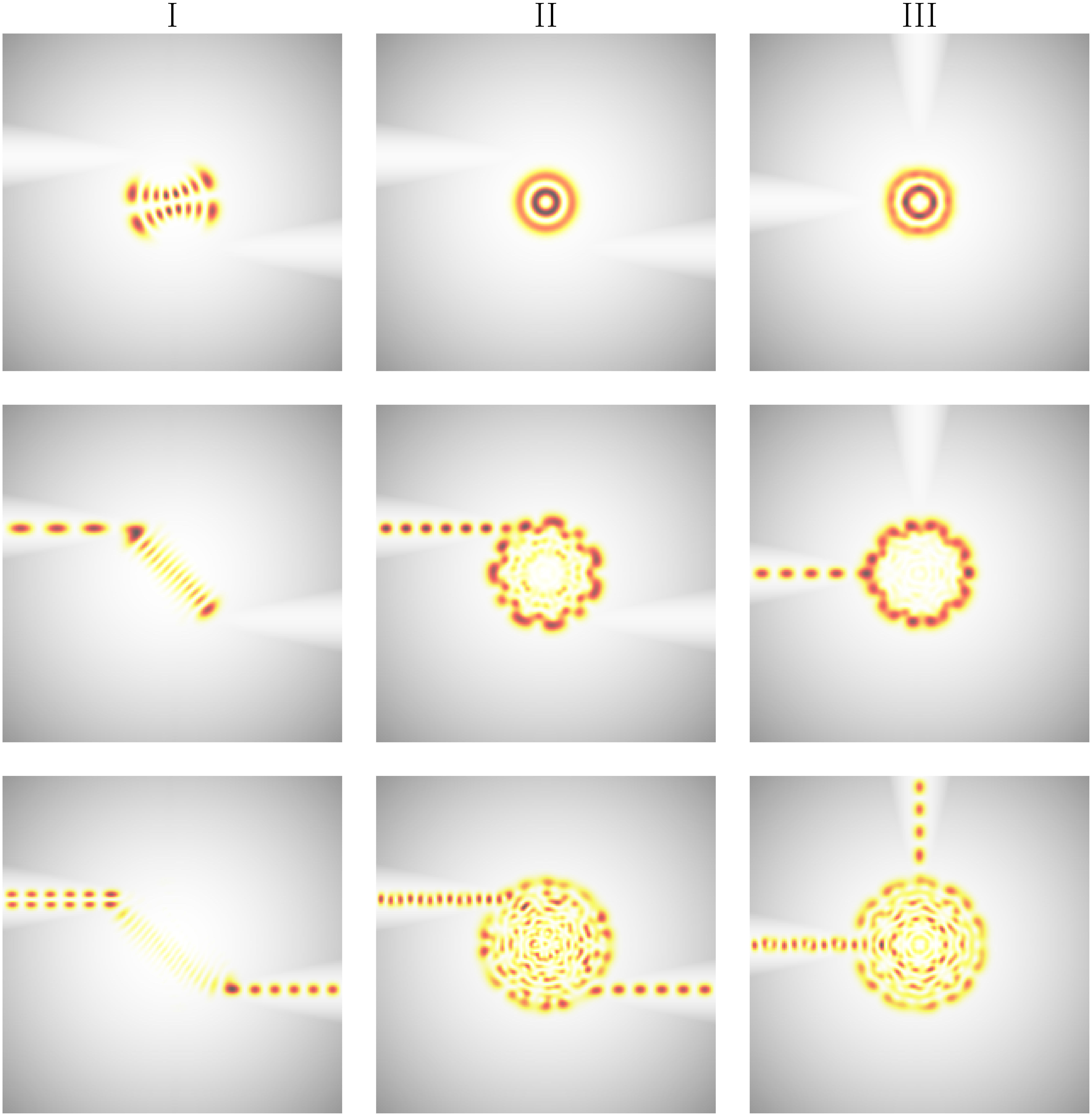}
    \caption{Examples of numerically solved eigenstates of systems I (left column), II (middle column), and III (right column). Eigenstates are selected in such a way, that their eigenenergies in the top, middle, and bottom rows are below $V_{0}$, between $V_{0}$ and $V_{1}$, and above $V_{1}$ respectively. The potential is superimposed on the eigenstate figures in greyscale to show the locations of the leads.}
    \label{fig:fockdarwinitp2d}
\end{figure}

The third case we consider is a 2D quantum dot system with a harmonic confining potential and strongly coupled leads. The system is exposed to a constant and uniform magnetic field perpendicular to the quantum dot plane. The Hamiltonian is written in a form
\begin{equation}
    \hat{H} = \hat{H}_{0} + \frac{1}{2}\omega^{2}(x^2+y^2) + \sum_{\alpha}V_{\alpha}(x,y),
\end{equation}
where $\hat{H}_{0}$ is the canonical Hamiltonian with magnetic field and $V_{\alpha}$ is the potential induced by the presence of the lead $\alpha$ in the system. In non-zero uniform magnetic fields this Hamiltonian corresponds to the well-known Fock-Darwin system, up to the inclusion of the potential terms associated with the leads. We consider three systems in zero temperature with varying magnetic field strengths $B$ and two-lead configurations:
\begin{itemize}
    \item System I: $B=0$, Leads 0 and 1 connecting to the quantum dot from left and right, respectively. 
    \item System II: $B=1$, Leads 0 and 1 connecting to the quantum dot from left and right, respectively.
    \item System III: $B=1$, Leads 0 and 1 connecting to the quantum dot from left and top, respectively.
\end{itemize}
We fix the bias energies of Leads 0 and 1 to be $V_{0}=10$ and $V_{1}=15$. The single-electron Schr{\"o}dinger equation for the central region is solved numerically with \textsc{itp2d}. Figure \ref{fig:fockdarwinitp2d} demonstrates some of the eigenstates. We note that below $V_{0}$ we do not observe any probability density "leaking" into the leads. As the eigenenergies surpass $V_{0}$, we start to observe probability density in Lead 0, until finally with eigenenergies higher than $V_{1}$ we see probability density in both Lead 0 and Lead 1. This is plausible behavior.

\begin{figure*}[ht]
    \centering
    \includegraphics[scale=0.16]{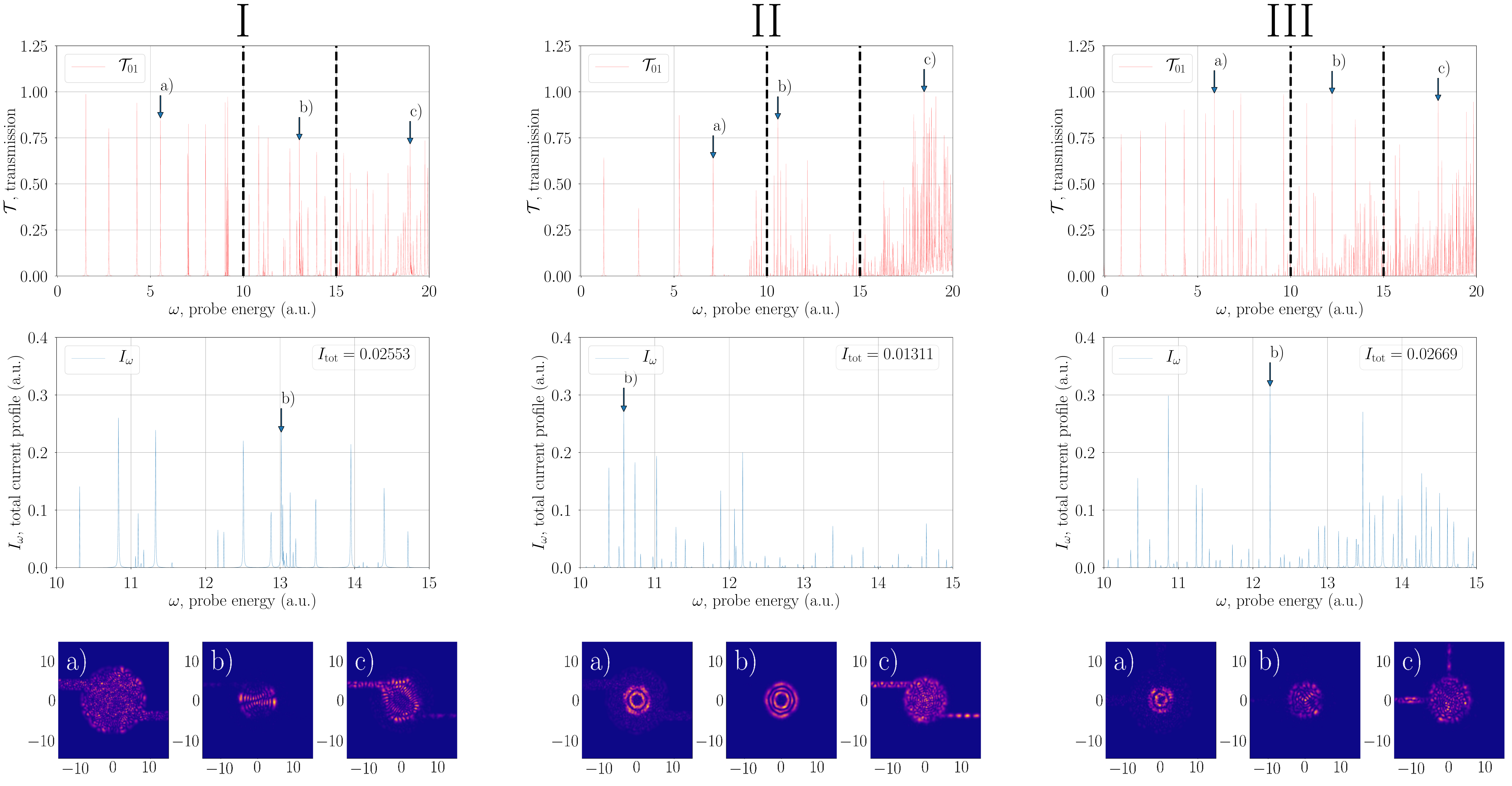}
    \caption{Transmissions (top row), $\omega$-dependent current profiles (middle row) and local density of states evaluated at a few select values of $\omega$ (bottom row) for systems I (left column), II (middle column), and III (right column). Only transmission from Lead 0 to Lead 1 has been included in the figure, as transmission from Lead 1 to Lead 0 had an identical shape. Similarly, as current profiles from Lead 0 to Lead 1 and from Lead 1 to Lead 0 are identical up to sign in case of a  two-lead system, only the positive profile is shown. Vertical dashed lines in transmission figures mark the interval between $V_{0}$ and $V_{1}$. Current profiles are displayed in the range between $V_{0}$ and $V_{1}$, as in zero temperature case there is current values outside of that range are zero. The figures showing the LDOS are normalized by their respective maximum values, and as such do not all share the same scale.}
    \label{fig:fockdarwintinie}
\end{figure*}

Next we utilize \textsc{tinie} to compute transmission, $\omega$-dependent current $I_{\omega}$ and total current running through each of systems I, II, and III specified above. Additionally, we computed the LDOS at a few probe energy values for each system: one corresponding to transmission peak below $V_{0}$, one corresponding to $\omega$-dependent current peak between $V_{0}$ and $V_{1}$, and one corresponding to transmission peak above $V_{1}$. Figure \ref{fig:fockdarwintinie} shows the results of the calculations.

We observe a complex structure of Dirac delta function-like peaks in both transmission and current profiles. The discrete nature of those peaks arises from the fact that our transport system has discrete energy levels both in the central region, as well as the leads. We can see that the number of peaks in transmission starts to increase drastically when $\omega > V_{1}$. At this point the electrons that are emitted from the lead regions have a sufficient amount of energy to propagate from one lead to another without getting confined in the central region. The LDOS in 
Figs.~\ref{fig:fockdarwintinie}(Ic), (IIc) and (IIIc) support this explanation. As we can see that there is state density present in both Lead 0 and Lead 1 regions.

We also find peaks below $V_{0}$ similarly to the potential barrier case in Sec. \ref{sec:barrier}. The peaks correspond to the electrons hitting the resonant energies. The LDOS gives us an insight into the nature of some of those resonant peaks. For instance, the LDOS in Figs.~\ref{fig:fockdarwintinie}(IIa) and (IIIa) resemble the eigenstates of the unperturbed system. Similarly, in the range between $V_{0}$ and $V_{1}$, we observe resonant peaks. 
In Figs.~\ref{fig:fockdarwintinie}(Ib) and (IIb) we once again see LDOS's that resemble the eigenstates of the unperturbed system, further providing evidence that some of the peaks are caused by the resonance of the states of the transport system with the eigenstates of the unperturbed system. However, not all the peaks can be explained in this manner. 
For example, the LDOS in Figs.~\ref{fig:fockdarwintinie}(Ia) and (IIb) do not resemble any unperturbed eigenstate; instead, they demonstrate complex nodal behavior. We point out that transmission from Lead 0 to Lead 1 is identical to that from Lead 1 to Lead 0, as it should be 
due to the conservation of the probability current.

The current profile is found to be non-zero only in the probe energy region between $V_{0}$ and $V_{1}$. This is to be expected in a zero-temperature case, as the difference between the Fermi-Dirac distributions of Eq.~(\ref{eq:pcurrents}) simplifies to a rectangular window. Moreover, we observe that the peaks in the current profile have the same locations as the peaks in the transmission within the considered energy range. This is due to Eq.~(\ref{eq:pcurrents}), which can be interpreted as the convolution over a rectangular window of the transmission. It preserves the peak locations in the window range, while removing those outside of it. The total current in Lead 0 is found to be the opposite of that for Lead 1, which makes sense due to the law of the current conservation.

\subsubsection{Multi-terminal two-dimensional perturbed potential well in a magnetic field}

As the final test case, we consider an extension of the previous example system by including soft boundaries and disorder. In particular, we study a 2D quantum dot modeled as a harmonic potential perturbed by local impurities under the influence of a perpendicular homogeneous magnetic field. The system has direct experimental relevance as a model for semiconductor quantum dots in the two-dimensional electron gas~\cite{reimann_rev.mod.phys_74_1283_2002, kouwenhoven_rep.prog.phys_64_701_2001}. Both theoretical and experimental studies have validated the harmonic approximation for modeling the external confinement of electrons in a quantum dot, even in the quantum Hall regime of high magnetic fields~\cite{rasanen_phys.rev.B_77_041302_2008, rogge_phys.rev.lett_105_046802_2010}. In addition, actual quantum devices such as quantum dots are affected by different types of impurities and imperfections (such as atoms or ions migrated into the system) that leave a signal in the measured magnetoconductance~\cite{halonen_phys.rev.B_53_6971_1996,hirose_phys.Rev.B_63_075301_2001, guclu_phys.rev.B_68_035304_2003,rasanen_phys.rev.B_70_115308_2004}. 
Furthermore, intentional perturbation of this kind can be generated in a controlled manner through a nanotip~\cite{bleszynski_nano.lett_7_2559_2007, boyd_nanotechnology_22_185201_2011, blasi_phys.rev.B_87_241303_2013}. It was recently discovered that disordered quantum wells display a new class of quantum scarring arising as a combined result of underlying near-degeneracies and localized perturbations ~\cite{Luukko_sci.rep_6_37656_2016, Keski-Rahkonen_phys.rev.b_96_094204_2017,Keski-Rahkonen_j.phys.:condens.matter_31_105301_2019, Keski-Rahkonen_phys.rev.lett_123_214101_2019}.
The counterintuitive nature of perturbation-induced scarring highlights the subtle role which disorder can play in quantum devices.

\begin{figure}[ht]
    \centering
    \includegraphics[scale=0.075]{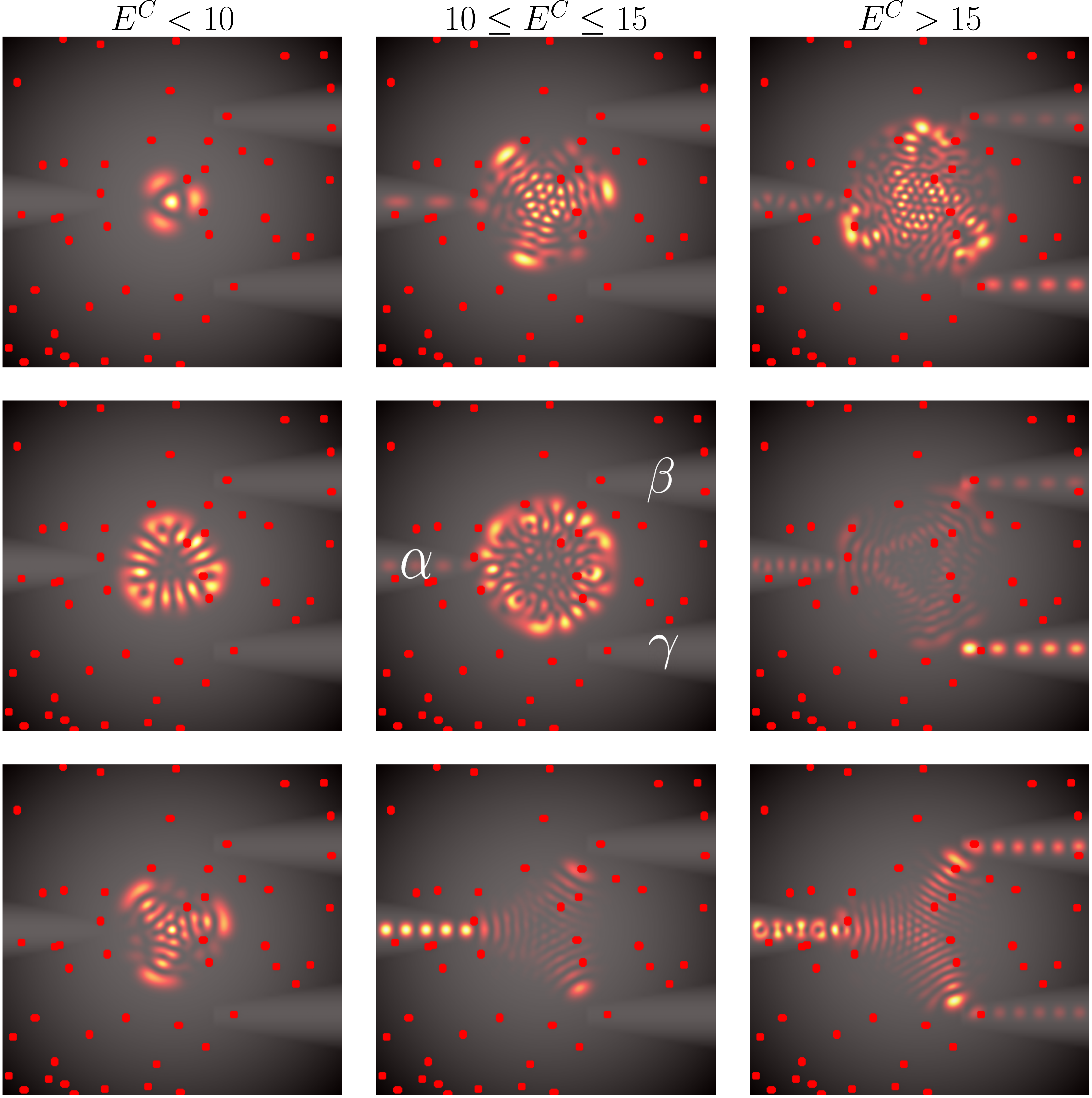}
    \caption{Examples of numerically solved eigenstates of the perturbed multi-lead 2D quantum dot. Each column provides a selection of three states within energy intervals specified on top. The confining potential is superimposed on the eigenstate figures to show the location of the leads. The locations of the Gaussian perturbations are marked with red dots.}
    \label{fig:perturbeditp2d}
\end{figure}

For testing purposes, we include additional complexity into the system by assuming that the modeled quantum dot is coupled to three leads, which are not at 90 degree angles with respect to each other (see labels $\alpha,\beta,$ and $\gamma$ in the central panel of Fig.~\ref{fig:perturbeditp2d}). The Hamiltonian is given by
\begin{equation}
    \hat{H} = \hat{H}_{0} + V_{\mathrm{ext}} + V_{\mathrm{imp}},
\end{equation}
where $\hat{H}_{0} = \frac{1}{2}\left(-i \nabla + \mathbf{A} \right)^2$ is the canonical Hamiltonian of a free electron. The perpendicular magnetic field $B$ is taken into account employing the linear gauge for the vector potential, i.e., $\mathbf{A}= (-B y, 0, 0)$. The external confining potential is given by
\begin{equation}
V_{\mathrm{ext}}(\mathbf{r}) = \mathrm{min}\left\{\frac{1}{2}\Vert \mathbf{r} \Vert^{2}, V_{\alpha}(\mathbf{r}),V_{\beta}(\mathbf{r}),V_{\gamma}(\mathbf{r})\right\}, 
\end{equation}
which contains a harmonic confinement in the central region as well as the lead potentials $V_{\alpha},V_{\beta},V_{\gamma}$ penetrating the system. 

The disorder is described as a sum of randomly distributed Gaussian bumps, i.e.,
\begin{equation}
    V_{\mathrm{imp}}(\mathbf{r}) = M \sum_{i} \exp \left[-\frac{\Vert \mathbf{r}-\mathbf{r}_{i} \Vert^2}{2\sigma^{2}}\right],
\end{equation}
where the sum goes over all randomized bump locations, and the individual bumps are defined by the amplitude $M=4$ and the width $\sigma=0.1$. 
We focus on the case where the bumps are randomly positioned with a uniform density of $0.1$ bumps per unit square. Thus there are many bumps inside the quantum device like shown in Fig.~\ref{fig:perturbeditp2d}. 
The system is also exposed to a perpendicular magnetic field of $B = 0.7$. 
 
The spatial confinement of the lead regions and the center region are defined as follows:
\begin{itemize}
    \item Center region: $x,y \in [-12,12]$;
    \item Lead $\alpha$: $x \in [-100,-3]$ and $y \in [-5,5]$;
    \item Lead $\beta$: $x \in [3,100]$ and $y \in [1,11]$;
    \item Lead $\gamma$: $x \in [3, 100]$ and $y \in [-11,-1]$.
\end{itemize}
The bias energies are set to $V_{\alpha}=10,V_{\beta}=15,$ and $V_{\gamma}=15$. We solve the single-electron Schrödinger equation numerically with \textsc{itp2d}. A selection of the eigenstates is presented in Figure~\ref{fig:perturbeditp2d}. The probability density is confined entirely to the quantum dot at eigenenrgies below $V_{\alpha}$, and starts leaking to Lead $\alpha$ as the eigenenergy surpasses the $V_{\alpha}$ threshold. Finally, when the eigenenergy surpasses the $V_{\beta}$ and $V_{\gamma}$, electron density is injeted 
through the center to Leads $\beta$ and $\gamma$ as expected. 


\begin{figure*}[ht]
    \centering
    \includegraphics[scale=0.16]{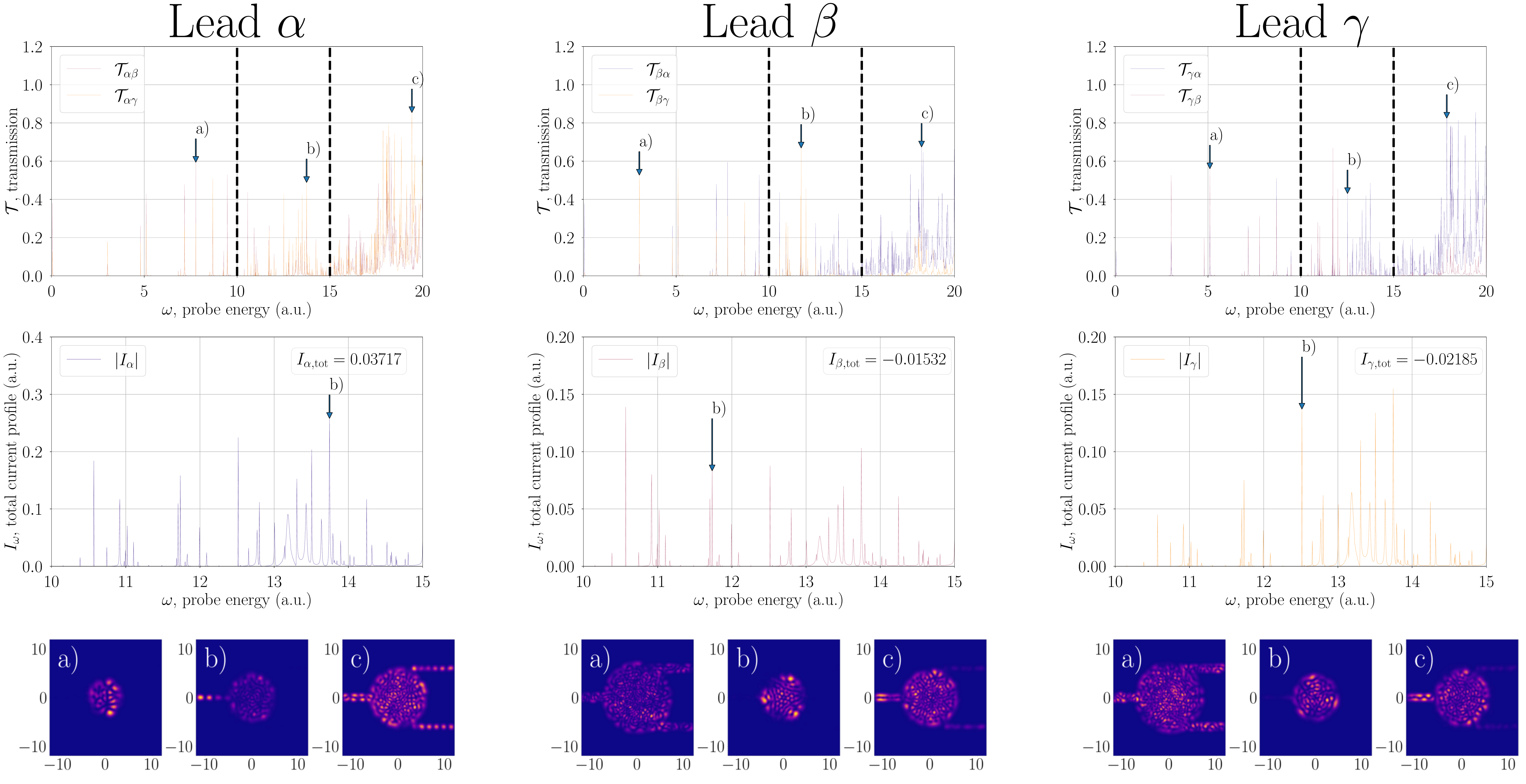}
    \caption{Transmissions (top row), $\omega$-dependent absolute current profiles (middle row) and local density of states (LDOS) evaluated at a few selected values of $\omega$ (bottom row) for the perturbed multi-lead 2D quantum dot. The columns separate transmission and absolute current profiles of Leads $\alpha, \beta,$ and $\gamma$ respectively. Vertical dashed lines in transmission mark the interval between $V_{\alpha}$ and $V_{\beta}=V_{\gamma}$. The current profiles are displayed in the range between $V_{\alpha}$ and $V_{\beta}=V_{\gamma}$. The figures showing the LDOS are normalized by their respective maximum values, and as such do not all share the same scale.}
    \label{fig:perturbedtinie}
\end{figure*}

Next we compute the transmission and current profiles in the probe energy range $\omega \in [0,20]$, as well as LDOS of a selection of states that yield high values of transmission. Specifically, for each of the leads we have selected three energy states that have high values of transmission: one below $V_{\alpha}$, one between $V_{\alpha}$ and $V_{\beta}$, and  one above $V_{\beta}$. Figure~\ref{fig:perturbedtinie} summarises the results of the calculations. Similarly to the test systems in Sec.~\ref{sec:fockdarwin}, we observe patterns of peaks that have a relatively low density below $V_{\alpha}$, but they become abundant at $[V_{\alpha},V_{\beta}]$ and especially so above $V_{\beta}$. In the LDOS we see the formation of the electronic states within the system that at first are localised within the quantum dot region. With increasing probe energies, LDOS starts to occupy Lead $\alpha$, and then finally Leads $\beta$ and $\gamma$ as well. The transmission profiles of the leads at the probe energy range $\omega \in [15,20]$ show that transmission favors electron flow between leads $\alpha$ and $\beta$, as well as between leads $\alpha$ and $\gamma$. The transmission coefficients are lower between leads $\beta$ and $\gamma$ as expected in view of the geometry of the system. 

Unlike the systems described in the preceding Subsections, some of the previously existing symmetries are broken because of the three-lead geometry and the applied magnetic field. For example, current profiles are not symmetric. Nonetheless, there are important principles that should be satisfied. First of all, the current through the quantum device has to be conserved, and indeed the total currents of the leads sum up to zero. This property is not explicitly enforced in \textsc{tinie}, but it arises as a consequence of the numerical scheme. Secondly, we have confirmed that the transmission functions obey the sum rules~\cite{Datta1997}. Finally, if we reverse the magnetic field and reverse the currents and voltage terminals, the conductance, or resistance, should be conserved. This reciprocity relation holds for a three-terminal conductance, which can be derived from the Landauer-Büttiker formalism (see, e.g., Ref.~\cite{Datta1997}), although the reciprocity property was originally derived for macroscopic conductors based on thermodynamical arguments. However, the Landauer-Büttiker formalism provides a terminal description in terms of measurable currents and voltages completely bypassing the details regarding the spatial variation of the potential inside a quantum device. We have confirmed that the reciprocity relations hold in our three-terminal perturbed quantum device. This further demonstrates that the computational framework is in solid agreement with the underlying theory of quantum transport.         

\subsection{Performance testing}

Here we present \textsc{tinie}'s performance benchmark results, which are based on timing the execution of \textsc{tinie} in the test cases of Sec. \ref{ssec:testcases}. The simulations have been performed in a HP Apollo 6000 XL230a Gen 9 supercluster with each node having two Intel Haswell E5-2690 v3 processors, i.e., 24 cores in a computing node. The results are summarized in Table \ref{tbl:performance}. 

\begin{table}[ht]
\renewcommand{\arraystretch}{1.25}
\caption{\textsc{tinie}'s test case performance benchmark. For one- and two-level systems, the number of states is computed as the number of non-zero elements in the rate operator matrix $\Gamma$ due to the use of WBLA in the computation. For the potential-barrier and Fock-Darwin systems, the number of states is computed as the sum of the numbers of elements in each of the coupling matrices in the system.}
\label{tbl:performance}
\begin{tabularx}{\linewidth}{c | c | c}
\hline
System & Number of States & Core time \\
\hline \hline
One-level $(T=0)$ & 1 & \SI{0.4}{core-sec} \\
One-level $(T=100)$ & 1 & \SI{5}{core-min} \\
Two-level & 2 & \SI{6}{core-min} \\
Potential barrier & $1.125\times10^8$ & \SI{7}{core-days} \\
Fock-Darwin & $6.4\times10^8$ & \SI{1}{core-month} \\
Perturbed 2D QD & $1.1883 \times 10^{8}$ & \SI{1.5}{core-days} \\
\hline
\end{tabularx}
\end{table}

Each test system has been evaluated at multiple values of the transport parameters, i.e., varying $\Gamma$ for one- and two-level systems and $T$ for the potential barrier system. A single \textsc{tinie} execution computes the transport properties of the system with one fixed set of transport parameters. In Table \ref{tbl:performance}, the times have been averaged over the total number of \textsc{tinie} executions performed during the test case computation. 

Additionally, the potential barrier system has been evaluated using 64 computing cores, leading to a wall time of 12 hours, further demonstrating the efficiency of the employed parallelization routines. Similarly, for the Fock-Darwin system, 32 computational cores have been used for \textsc{tinie\_prepare} stage, and 16 cores have been used for \textsc{tinie} stage of the computation. This results in overall wall time of 41 hours. Faster execution time of the final test system comes from utilizing a reduced resolution of the eigenfunction spatial grid during the \textsc{tinie\_prepare} stage, which proved to still be sufficiently accurate. These results confirm that \textsc{tinie} is can indeed perform \textit{ab initio} transport calculations within a reasonable time frame.


\section{Summary}
\label{sec:summary}

We have presented \textsc{tinie} -- a computational simulation framework for quantum transport in two-dimensional systems of arbitrary geometry. \textsc{tinie} Python package provides a comprehensive toolset for quantum transport phenomena in nanoscale systems. \textsc{tinie} performs its transport calculations from the first principles, that is, employing the Landauer-Büttiker approach, without any approximations besides those related to the wave function discretization and numerical differentiation/integration. One of \textsc{tinie}'s core strengths is its capability to perform calculations in a reasonable time without the need to resort to approximate the theoretical transport formalism. However, \textsc{tinie} is also capable of utilizing some of the approximation regimes, such as the wide band limit approximation. The modular structure of \textsc{tinie} allows for easy expansion and compatibility with external software, such as \textsc{itp2d} Schr{\"o}dinger equation solver. This enables to use the optimized algorithms geared for specific problems, and an expansion to easily implement different kinds of coupling schemes. 

In this paper, we have demonstrated \textsc{tinie}'s versatility has been demonstrated in various test cases. We have investigated simple one- and two-state systems, a potential barrier system, and a realistic two-dimensional quantum dot system in a magnetic field without and with random Gaussian impurities. Each test case produces reasonable results of high numerical accuracy. Hence, the examples demonstrate \textsc{tinie}'s suitability and flexibility for studying transport phenomena in two dimensions. In general, \textsc{tinie} can be used to gain specific information about the effects of a magnetic field and disorder in quantum devices. One near-future application for \textsc{tinie} is to study the effect of perturbation-induced scars on quantum transport in disordered nanostructures~\cite{Luukko_sci.rep_6_37656_2016, Keski-Rahkonen_phys.rev.b_96_094204_2017,Keski-Rahkonen_j.phys.:condens.matter_31_105301_2019, Keski-Rahkonen_phys.rev.lett_123_214101_2019}.

\section*{Acknowledgments}

We are grateful to Tobias Kramer, Henri Saarikoski, and Matti Molkkari for fruitful collaboration and discussions. We also acknowledge CSC -- Finnish IT Center for Science for computational resources. Furthermore, J. K-R. thanks the Emil Aaltonen Foundation for financial support.





\bibliographystyle{elsarticle-num}
\bibliography{references.bib}







\end{document}